\documentclass[a4paper,fleqn]{cas-dc}
\pdfoutput=1
\usepackage[numbers,sort&compress]{natbib}
\usepackage[T1]{fontenc}
\usepackage[utf8]{inputenc}
\usepackage[english]{babel}
\usepackage[protrusion=true,expansion=false]{microtype}
\usepackage{amsmath,amssymb,amsfonts,amsbsy,mathtools}
\usepackage{multirow,dcolumn,rotating,overpic,float,bbold,enumerate,dsfont,mathrsfs,tensor}
\usepackage{caption,subcaption,ragged2e,booktabs}
\usepackage{placeins}
\usepackage[dvipsnames]{xcolor}
\usepackage{orcidlink}
\usepackage[normalem]{ulem} 
\usepackage{newunicodechar}
\newunicodechar{Λ}{$\Lambda$}
\definecolor{palatinateblue}{rgb}{0.15,0.23,0.89}
\definecolor{amaranth}{rgb}{0.90,0.17,0.31}
\definecolor{brightpink}{rgb}{0.86,0.26,0.55}

\hypersetup{colorlinks=true,linkcolor=palatinateblue,citecolor=palatinateblue,urlcolor=palatinateblue,
 pdftitle={Three-form dark energy: constraints and multi-probe comparison with $\Lambda$CDM},
 pdfauthor={Mariam Bouhmadi-Lopez; Hsu-Wen Chiang; Carlos G. Boiza; Javier Ortega del Rio; Thomas J. Broadhurst; Pisin Chen}}

\newcommand{\AuthorMetaLinks}[2]{%
  \orcidlink{#2}%
}
\newcommand{\EmailTextLink}[1]{\href{mailto:#1}{\ttfamily #1}}

\newcommand{\AffiliationTarget}[1]{%
  \cspreto{ca_affiliation_#1}{\hypertarget{aff:#1}{}}%
}

\ExplSyntaxOn
\RenewDocumentCommand \processAffNum { m }
  {
    \seq_clear:N \l_affmk_seq
    \clist_map_inline:nn { #1 }
      {
        \seq_put_right:Nn \l_affmk_seq
          { \hyperlink{aff:##1}{\int_to_alph:n {##1}} }
      }
  }

\RenewDocumentCommand \emailauthor { m m }
  {
    \int_gincr:N \g_ead_int
    \str_case:nnF {#1}
      {
        {mariam.bouhmadi@ehu.eus}
          {\seq_gput_right:Nn \g_stm_ead_seq {\mbox{\EmailTextLink{mariam.bouhmadi@ehu.eus}\ (M.~Bouhmadi-L\'opez)}}}
        {jiangxw@sustech.edu.cn}
          {\seq_gput_right:Nn \g_stm_ead_seq {\mbox{\EmailTextLink{jiangxw@sustech.edu.cn}\ (H.~Chiang)}}}
        {b98202036@ntu.edu.tw}
          {\seq_gput_right:Nn \g_stm_ead_seq {\mbox{\EmailTextLink{b98202036@ntu.edu.tw}\ (H.~Chiang)}}}
        {carlos.garciab@ehu.eus}
          {\seq_gput_right:Nn \g_stm_ead_seq {\mbox{\EmailTextLink{carlos.garciab@ehu.eus}\ (C.G.~Boiza)}}}
        {javier.ortega@ehu.eus}
          {\seq_gput_right:Nn \g_stm_ead_seq {\mbox{\EmailTextLink{javier.ortega@ehu.eus}\ (J.O.d.~R\'io)}}}
        {tom.j.broadhurst@gmail.com}
          {\seq_gput_right:Nn \g_stm_ead_seq {\mbox{\EmailTextLink{tom.j.broadhurst@gmail.com}\ (T.J.~Broadhurst)}}}
        {pisinchen@phys.ntu.edu.tw}
          {\seq_gput_right:Nn \g_stm_ead_seq {\mbox{\EmailTextLink{pisinchen@phys.ntu.edu.tw}\ (P.~Chen)}}}
      }
      {\seq_gput_right:Nn \g_stm_ead_seq {\mbox{\EmailTextLink{#1}}}}
  }

\RenewDocumentCommand \printemails { }
  {
    \group_begin:
    \int_compare:nNnTF { \int_use:N \g_ead_int } > { 0 }
      {
        \tex_let:D \thefootnote \relax \footnotetext
          {
            \raggedright
            \textit{Email~addresses:}\par\nobreak\smallskip
            \seq_use:Nn \g_stm_ead_seq { ;\space }
          }
      }
      { }
    \group_end:
  }

\RenewDocumentCommand \printorcid { } { }
\ExplSyntaxOff

\begin{document}
\let\WriteBookmarks\relax
\def\floatpagepagefraction{0.55}
\def\textpagefraction{0.001}
\renewcommand{\floatpagefraction}{0.55}
\renewcommand{\dblfloatpagefraction}{0.55}
\raggedbottom

\shorttitle{}
\shortauthors{Bouhmadi-López et al.}

\ExplSyntaxOn
\cs_set:Npn \__first_footerline: {}
\ExplSyntaxOff

\title[mode=title]{Three-form dark energy: constraints and multi-probe comparison with $\Lambda$CDM}

\author[1,2,3]{Mariam Bouhmadi-López}[suffix={\AuthorMetaLinks{mariam.bouhmadi@ehu.eus}{0000-0002-1529-1889}}]
\ead{mariam.bouhmadi@ehu.eus}
\author[4]{Hsu-Wen Chiang}[suffix={\AuthorMetaLinks{jiangxw@sustech.edu.cn}{0000-0002-5450-9297}}]
\ead{jiangxw@sustech.edu.cn}
\ead{b98202036@ntu.edu.tw}
\author[2,3]{Carlos G. Boiza}[suffix={\AuthorMetaLinks{carlos.garciab@ehu.eus}{0000-0002-1009-8115}}]
\ead{carlos.garciab@ehu.eus}
\author[2,3]{\linebreak Javier Ortega del Río}[suffix={\AuthorMetaLinks{javier.ortega@ehu.eus}{0009-0005-5887-0556}}]
\cormark[1]
\ead{javier.ortega@ehu.eus}
\author[1,2,3,5]{Thomas J. Broadhurst}[suffix={\AuthorMetaLinks{tom.j.broadhurst@gmail.com}{0000-0002-8785-8979}}]
\ead{tom.j.broadhurst@gmail.com}
\author[6,7]{Pisin Chen}[suffix={\AuthorMetaLinks{pisinchen@phys.ntu.edu.tw}{0000-0001-5251-7210}}]
\ead{pisinchen@phys.ntu.edu.tw}

\cortext[1]{Corresponding author}

\affiliation[1]{organization={IKERBASQUE, Basque Foundation for Science},
                 postcode={48011}, city={Bilbao}, country={Spain}}
\AffiliationTarget{1}
\affiliation[2]{organization={Department of Physics, University of the Basque Country UPV/EHU},
                 addressline={P.O. Box 644}, postcode={48080}, city={Bilbao}, country={Spain}}
\AffiliationTarget{2}
\affiliation[3]{organization={EHU Quantum Center, University of the Basque Country UPV/EHU},
                 addressline={P.O. Box 644}, postcode={48080}, city={Bilbao}, country={Spain}}
\AffiliationTarget{3}
\affiliation[4]{organization={Department of Physics, Southern University of Science and Technology},
                 postcode={518055}, city={Shenzhen}, country={China}}
\AffiliationTarget{4}
\affiliation[5]{organization={Donostia International Physics Center},
                 addressline={Paseo Manuel de Lardizabal, 4}, postcode={20018}, city={San Sebastián}, country={Spain}}
\AffiliationTarget{5}
\affiliation[6]{organization={Leung Center for Cosmology and Particle Astrophysics, National Taiwan University},
                 postcode={10617}, city={Taipei}, country={Taiwan}}
\AffiliationTarget{6}
\affiliation[7]{organization={Department of Physics and Graduate Institute of Astrophysics, National Taiwan University},
                 postcode={10617}, city={Taipei}, country={Taiwan}}
\AffiliationTarget{7}


\begin{abstract}
Three-form fields provide a theoretically well-motivated framework for dark energy, arising in higher-dimensional theories and exhibiting a rich cosmological phenomenology. We investigate a minimally coupled three-form dark energy model with a Gaussian potential and constrain it using current cosmological observations, including CMB shift parameters, DESI DR2 baryon acoustic oscillation measurements, Pantheon+ supernovae with and without SH0ES calibration, cosmic chronometers, and gamma-ray bursts. Parameter estimation is performed within a Bayesian Markov-chain Monte Carlo framework, while model comparison relies on several information criteria and the Bayesian evidence, as well as tension statistics. We find that the three-form model provides a viable and competitive description of the expansion history.
It is mildly preferred over $\Lambda$CDM for the combination of early and late-time datasets that are heavily tensioned (CMB+BAO and Pantheon+SH0ES). This preference decreases to neutrality for the other, less tensioned combination of early-time and late-time data, while for individual early-time or late-time datasets analysed separately, the information criteria are neutral or favour $\Lambda$CDM. This suggests that the additional degrees of freedom associated with the three-form field may help accommodate cosmological observations of different origins within a common framework. The reconstructed dark energy dynamics exhibit a phantom phase at intermediate redshifts while approaching a cosmological-constant-like behaviour at early and late times, providing a distinctive observational signature. Although the model does not significantly alleviate the Hubble tension despite allowing higher values of $H_0$, it remains consistent with current observations and offers a well-motivated alternative to $\Lambda$CDM whose predictions can be tested by future cosmological surveys.
\end{abstract}


\maketitle

\section{Introduction}\label{intro}

A wide range of cosmological observations has established that the expansion of the Universe is currently accelerating, indicating the presence of a dominant component with negative effective pressure at late times \cite{SupernovaSearchTeam:1998fmf,SupernovaCosmologyProject:1998vns}. Within the standard $\Lambda$CDM cosmological model, this phenomenon is attributed to a cosmological constant, which provides a remarkably successful description of the cosmic expansion history and the formation of large-scale structure when confronted with data from the cosmic microwave background, baryon acoustic oscillations, and supernova observations \cite{Planck2018}. Nevertheless, when different observational probes are analysed jointly, mild but persistent inconsistencies emerge, most notably in the determination of the present day Hubble parameter. This so called Hubble tension reflects a discrepancy between early Universe inferences obtained within the $\Lambda$CDM framework and late-Universe distance-ladder measurements of $H_0$ \cite{Riess:2019cxk,Verde:2019ivm}, and has prompted growing interest in extensions of the standard cosmological model that modify the late-time expansion history \cite{CosmoVerseNetwork:2025alb,DiValentino:2026uua,Cai:2026swf}.

In addition to this observational tension, the cosmological constant also faces significant theoretical challenges. Interpreting $\Lambda$ as vacuum energy leads to a severe mismatch between its observed value and theoretical expectations from quantum field theory, commonly known as the fine-tuning problem \cite{Weinberg:1988cp}. Moreover, the coincidence problem raises the question of why the energy densities of matter and dark energy are of comparable magnitude precisely at the present epoch, despite their distinct evolutionary behaviour \cite{Padmanabhan:2002ji}. Together, these observational and theoretical considerations have motivated the exploration of dynamical dark energy models and alternative theories of gravity as possible extensions of the standard cosmological framework \cite{CANTATA:2021asi}.

Motivated by both theoretical considerations and observational challenges, a wide variety of extensions of the standard $\Lambda$CDM model have been proposed to account for the origin of cosmic acceleration \cite{Capozziello:2011et, DiValentino2021}. One prominent class of scenarios describes dark energy as a dynamical scalar field, such as quintessence or its generalisations, whose evolving equation of state can drive late-time accelerated expansion and potentially alleviate conceptual issues associated with a constant cosmological constant \cite{Ratra:1987rm,Caldwell1998,Copeland:2006wr,Boiza:2024azh,Chiang:2025qxg}. An alternative, more phenomenological approach models dark energy as an effective perfect fluid with a parametrised equation of state, offering a flexible framework to explore departures from $\Lambda$CDM at the background level \cite{Chevallier:2000qy,Linder:2002et,Kamenshchik:2001cp,Bento:2002ps,Bouhmadi-Lopez:2006fwq,Bouhmadi-Lopez:2014cca}. 

A conceptually distinct direction attributes the observed acceleration to modifications of General Relativity on cosmological scales, where the late-time dynamics arise from extended gravitational interactions rather than from an additional energy component \cite{Capozziello:2011et,Nojiri:2017ncd}. Well-studied examples of modified-gravity frameworks include extensions of the Einstein-Hilbert action, such as metric and Palatini formulations in which the gravitational action generalises the linear dependence on the scalar curvature \cite{Capozziello:2011et,Nojiri:2010wj,Nojiri:2017ncd,Morais:2015ooa}, teleparallel formulations constructed from the torsion scalar or from general functions thereof \cite{Bengochea:2008gz,Ferraro:2006jd,Cai:2015emx}, as well as more recent frameworks based on spacetime non-metricity, where the gravitational action is built from scalar invariants of the non-metricity tensor \cite{BeltranJimenez:2018vdo,BeltranJimenez:2019tme,Ayuso:2020dcu,Boiza:2025xpn,Ayuso:2025vkc}.

More recently, scenarios in which the effective dark energy density or cosmological term changes sign during cosmic evolution have attracted increasing attention, as they offer novel mechanisms to address the coincidence problem and late-time tensions within a unified framework \cite{Akarsu:2021fol,Akarsu:2022typ,Akarsu:2023mfb,Akarsu:2025nns,Bouhmadi-Lopez:2025ggl,Bouhmadi-Lopez:2025spo,Ibarra-Uriondo:2026zbp,Bouhmadi-Lopez:2026vyc}. Taken together, these complementary approaches provide a broad theoretical landscape for investigating the physics underlying cosmic acceleration and motivate systematic observational tests of departures from the simplest cosmological paradigm.

On the other hand, three-form fields have been extensively studied in cosmological model building as viable ingredients for both early-time accelerated expansion and late-time dark energy \cite{Koivisto:2009fb,Koivisto_2013,Morais:2016bev,Bouhmadi-Lopez:2016dzw,Morais:2017vlf,Klinkhamer:2007pe,Klinkhamer:2008ns}. Such fields arise naturally in higher-dimensional theories, including string theory and supergravity \cite{Aurilia:1980xj}, and can generate an effective negative pressure capable of driving accelerated expansion \cite{Koivisto:2009ew}. For specific choices of the potential, including constant or vanishing forms, their background evolution can mimic that of a cosmological constant \cite{Duff:1980qv}, allowing for a unified description of inflation or late-time cosmic acceleration within the same framework \cite{Koivisto:2009fb,Morais:2016bev}.

At late times, three-form fields have been extensively studied as dark energy candidates, exhibiting de Sitter attractors, quintessence-like dynamics, or phantom behaviour depending on the potential and possible interactions with dark matter \cite{Morais:2016bev,Koivisto:2009fb,Koivisto_2013,Morais:2017vlf}. Phantom regimes further allow for the construction of wormholes \cite{Bouhmadi-Lopez:2021zwt,Barros_2018}, regular black holes \cite{Bouhmadi-Lopez:2020wve,Barros:2020ghz}, black-bounce geometries \cite{Lobo:2026flk}, and compact stellar configurations \cite{Barros:2021jbt}. Beyond late-time cosmology, three-form fields have also been explored in inflationary single- and multi-field scenarios \cite{Koivisto:2009fb,Koivisto:2009ew,Germani:2009iq,Mulryne_2012,Urban:2012ib,Kumar:2014oka,SravanKumar:2016biw}, with studies of stability, reheating, and phenomenology \cite{DeFelice:2012jt,DeFelice:2012wy}, as well as extensions to brane-world models \cite{Barros:2015evi,Barros:2023nzr,Gordin:2023nsv}. Further developments include k-essence extensions \cite{daFonseca:2024boz}, anisotropic cosmologies \cite{DeFelice:2025khe}, screening mechanisms compatible with Solar System tests \cite{Barreiro:2016aln}, and quantum cosmology \cite{Bouhmadi-Lopez:2018lly}.

Despite the breadth of existing theoretical studies, observational constraints on dark energy models driven by three-form fields remain relatively scarce. To the best of our knowledge, the first dedicated observational analysis of such models was presented in Ref.~\cite{Bouhmadi-Lopez:2025lzm}. The primary aim of the present work is to extend this line of investigation by deriving new observational constraints and by examining whether the phantom-like behaviour that naturally emerges in three-form dark energy scenarios can play a role in alleviating current cosmological tensions. In contrast to our previous study \cite{Bouhmadi-Lopez:2025lzm}, which incorporated perturbative observables, the analysis presented here focuses predominantly on background-level cosmological data, thereby providing a complementary assessment of the phenomenological viability of these models.

The remainder of this paper is organised as follows. In Section~\ref{sec:gen_three-form_summary},
we present the theoretical framework underlying our analysis and provide a concise summary of
the three-form field model under consideration. Section~\ref{observations} introduces the
cosmological datasets used in this analysis: CMB shift parameters, baryon acoustic oscillations, type Ia supernovae (with and without the SH0ES
distance-ladder calibration), cosmic chronometers, and gamma-ray bursts.
Section~\ref{sec:methodology} describes the statistical methodology used to confront the model
with these data, including the likelihoods, parameter priors, and model-comparison criteria
adopted in the analysis. Section~\ref{fits} presents the resulting parameter constraints, as well as the model-comparison and
tension-analysis results across the dataset combinations considered. Finally,
Section~\ref{conclusions} summarises the main findings of this study and offers concluding
remarks and perspectives for future work.

\section{Three-form dark energy: theoretical framework and dynamics}
\label{sec:gen_three-form_summary}

We briefly summarise the theoretical framework and dynamical properties of the three-form dark energy model introduced in Refs.~\cite{Koivisto:2009fb,Morais:2016bev}, which forms the basis of the present analysis. The model consists of a minimally coupled three-form field $A_{\mu\nu\rho}$ with a self-interaction potential $V(A^2)$, evolving in a spatially flat Friedmann-Lemaître-Robertson-Walker (FLRW) Universe filled with standard radiation and pressureless matter. The dynamics of the system are derived from the action
\begin{align}
S ={}& \int \mathrm{d}^4x \sqrt{-g}\left[
\frac{1}{2\kappa^2} R
- \frac{1}{48} F_{\mu\nu\rho\sigma}F^{\mu\nu\rho\sigma}
\right. \nonumber\\
&\left.{}- V(A^2)\right] + S_{\rm r} + S_{\rm m},
\end{align}
where $\kappa^2 = 8\pi G$, $F_{\mu\nu\rho\sigma} = \nabla_{[\mu} A_{\nu\rho\sigma]}$ denotes the associated four-form field strength, and $A^2 \equiv A_{\mu\nu\rho}A^{\mu\nu\rho}$. The terms $S_{\rm r}$ and $S_{\rm m}$ represent the actions describing the radiation and pressureless matter components, respectively.

Assuming homogeneity and isotropy, the three-form field can be parametrised as
\begin{equation}
A_{ijk} = a^3(t)\,\epsilon_{ijk}\,\chi(t),
\end{equation}
where $a(t)$ is the scale factor, $\epsilon_{ijk}$ is the Levi-Civita symbol, and $\chi(t)$ denotes the effective scalar degree of freedom associated with the three-form. In this background, the equation of motion for $\chi$ takes the form
\begin{equation}
\ddot{\chi} + 3H\dot{\chi} + 3\dot{H}\chi + V_{,\chi} = 0,
\end{equation}
where $H \equiv \dot{a}/a$ and $V_{,\chi} \equiv \mathrm{d}V/\mathrm{d}\chi$.

The effective energy density and pressure associated with the three-form field are given by
\begin{eqnarray}\label{rho&pX}
\rho_\chi &=& \frac{1}{2}\left(\dot{\chi} + 3H\chi\right)^2 + V(\chi),\nonumber \\
p_\chi &=& -\frac{1}{2}\left(\dot{\chi} + 3H\chi\right)^2 - V(\chi) + V_{,\chi}\chi .
\end{eqnarray}
The corresponding equation-of-state parameter of the three-form field is then defined as
\begin{equation}\label{EoSthree-form}
w_\chi \equiv \frac{p_\chi}{\rho_\chi}
= -1 + \frac{V_{,\chi}\chi}{
\frac{1}{2}\left(\dot{\chi} + 3H\chi\right)^2 + V(\chi)
}.
\end{equation}
These quantities govern the background cosmological evolution through the Friedmann equations,
\begin{align}
H^2 &= \frac{\kappa^2}{3}\left(\rho_{\rm r} + \rho_{\rm m} + \rho_\chi\right), \\
\frac{\ddot{a}}{a} &= -\frac{\kappa^2}{6}
\left(\rho_{\rm m} + 2\rho_{\rm r} + \rho_\chi + 3p_\chi\right),
\end{align}
while the radiation and matter components are assumed to be minimally coupled and independently conserved, satisfying the standard continuity equations,
\begin{equation}
\dot{\rho}_{\rm r} + 4H\rho_{\rm r} = 0,
\qquad
\dot{\rho}_{\rm m} + 3H\rho_{\rm m} = 0.
\end{equation}

Depending on the form of the potential $V(\chi)$, the three-form field can exhibit a wide variety of cosmological behaviours (see Eq.~(\ref{EoSthree-form})). In particular, constant or vanishing potentials reproduce an effective cosmological constant, whereas more general choices lead to genuinely dynamical dark energy evolution. 

In this work, we focus on a positive Gaussian potential as a concrete and well-motivated realisation capable of driving late-time cosmic acceleration. Following the notation of Ref.~\cite{Bouhmadi-Lopez:2025lzm}, the three-form potential is taken to be
\begin{equation}
V(\chi)=V_\ast\,\exp\!\left(-\frac{\xi\kappa^2}{6}\chi^2\right),
\qquad V_\ast,\xi>0,
\end{equation}
where $V_\ast$ sets the overall energy scale of the potential and $\xi$ is a dimensionless parameter controlling the width of the Gaussian potential. Since $V_{,\chi}\chi<0$ for $\chi\neq0$, Eq.~(\ref{EoSthree-form}) implies that the three-form generically exhibits phantom-like behaviour once the field is displaced from the maximum at $\chi=0$.

\paragraph{Effective potential and late-time dynamics.}

Starting from the background equation of motion
\begin{equation}
\ddot{\chi}+3H\dot{\chi}+3\dot{H}\chi+V_{,\chi}=0,
\label{eq:chi_eom_basic}
\end{equation}
we eliminate $\dot H$ using the Raychaudhuri equation for a spatially flat FLRW Universe,
\begin{equation}
\dot{H}=-\frac{\kappa^2}{2}\left(\rho_{\rm tot}+p_{\rm tot}\right)
=-\frac{\kappa^2}{2}\left(\rho_{\rm m}+\frac{4}{3}\rho_{\rm r}+\rho_\chi+p_\chi\right).
\label{eq:raychaudhuri}
\end{equation}
For a minimally coupled three-form, Eq.~\eqref{rho&pX} implies
\begin{equation}
\rho_\chi+p_\chi = V_{,\chi}\chi,
\label{eq:rho_plus_p_3form}
\end{equation}
so that
\begin{equation}
\dot{H}=-\frac{\kappa^2}{2}\left(\rho_{\rm m}+\frac{4}{3}\rho_{\rm r}+V_{,\chi}\chi\right).
\label{eq:Hdot_explicit}
\end{equation}
Substituting Eq.~(\ref{eq:Hdot_explicit}) into Eq.~(\ref{eq:chi_eom_basic}) yields
\begin{equation}
\ddot{\chi}+3H\dot{\chi}
+\left(1-\frac{\chi^2}{\chi_c^2}\right)V_{,\chi}
=
\frac{3\kappa^2}{2}\left(\rho_{\rm m}+\frac{4}{3}\rho_{\rm r}\right)\chi,
\label{eq:chi_eom_ray}
\end{equation}
where we have introduced the critical field amplitude \cite{Morais:2016bev,Bouhmadi-Lopez:2016dzw}
\begin{equation}
\chi_c \equiv \sqrt{\frac{2}{3\kappa^2}}.
\label{eq:chi_c_def}
\end{equation}
In the late-time regime, when matter and radiation dilute away, the evolution is governed by
\begin{equation}
\ddot{\chi}+3H\dot{\chi}
+\left(1-\frac{\chi^2}{\chi_c^2}\right)V_{,\chi}=0.
\label{eq:chi_eom_late}
\end{equation}

This motivates defining an effective potential through
\begin{equation}
\frac{\partial V_{\rm eff}}{\partial \chi}
=
\left(1-\frac{\chi^2}{\chi_c^2}\right)V_{,\chi}.
\label{Veff_def_full}
\end{equation}
Stationary configurations therefore satisfy either $V_{,\chi}=0$ or $\chi=\pm\chi_c$. Thus,
besides the maximum of the Gaussian potential at $\chi=0$, the effective force introduces dynamically
distinguished field amplitudes at $\chi=\pm\chi_c$, which organise the global phase-space structure and
are directly related to the existence of the Little Sibling of the Big Rip (LSBR)\footnote{The little sibling of the big rip (LSBR) corresponds to a future asymptotic regime in which the scale factor and the Hubble parameter diverge only at infinite cosmic time, while $\dot H$ remains finite, leading to the eventual disruption of bound structures without the occurrence of a true spacetime singularity. As a consequence, the LSBR is reached only at infinite cosmic time and does not correspond to a spacetime singularity or geodesic incompleteness. Nevertheless, the unbounded growth of the Hubble parameter leads to a diverging inertial force, which eventually causes the dissociation of gravitationally bound structures. The LSBR therefore represents an intermediate fate of the Universe, lying between smooth phantom acceleration and the more violent Big Rip singularity.}
attractor \cite{Bouhmadi-Lopez:2014cca,Morais:2016bev}.

For the Gaussian potential, defining
\begin{equation}
\alpha\equiv\frac{\xi\kappa^2}{6},
\label{eq:a_def}
\end{equation}
one finds explicitly, up to an additive constant,
\begin{equation}
V_{\rm eff}(\chi)
=
V(\chi)\left[
1-\frac{\chi^2}{\chi_c^2}
-\frac{1}{\alpha\,\chi_c^2}
\right]
+ C.
\label{Veff_explicit_full}
\end{equation}
We fix the constant $C$ by imposing
$V_{\rm eff}(\chi_c)=0$, which gives
\begin{equation}
C=\frac{V(\chi_c)}{\alpha\,\chi_c^2}.
\label{eq:C_choice_general}
\end{equation}
Using Eqs.~(\ref{eq:chi_c_def}) and (\ref{eq:a_def}), this can be written explicitly as
\begin{equation}
\alpha\,\chi_c^2=\frac{\xi}{9},
\quad
V(\chi_c)=V_\ast\,e^{-\xi/9},
\quad\Rightarrow\quad
C=\frac{9}{\xi}\,V_\ast\,e^{-\xi/9}.
\label{eq:C_choice_explicit}
\end{equation}

\begin{figure}
\centering
\includegraphics[width=0.95\columnwidth]{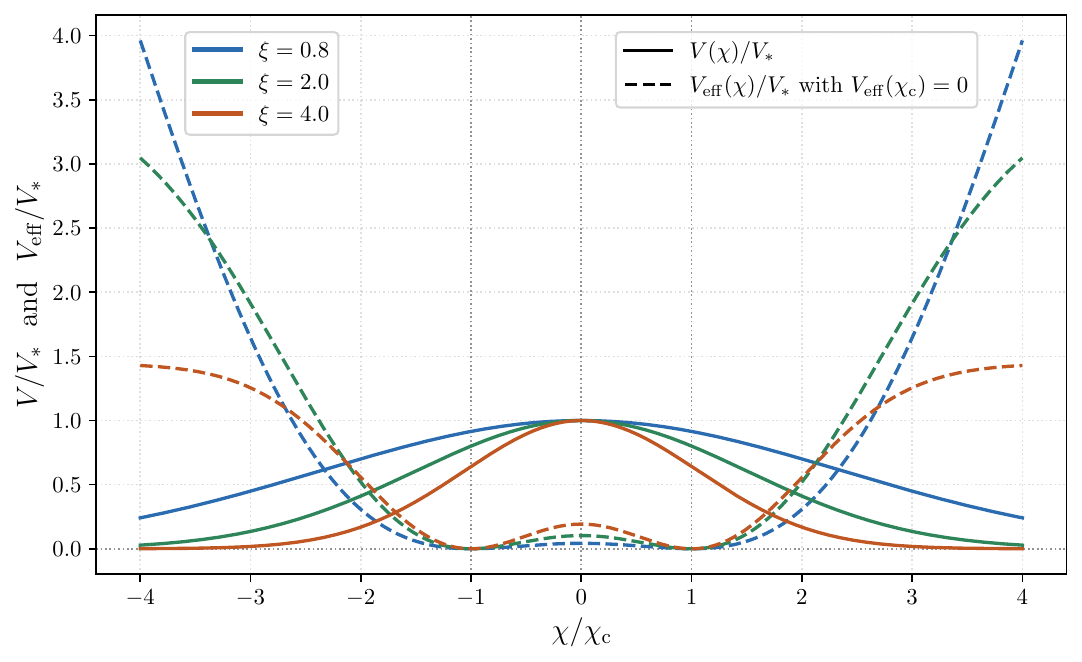}
\caption{\justifying{Gaussian three-form potential $V(\chi)$ and the corresponding effective potential $V_{\rm eff}(\chi)$ for representative parameter values.}}
\label{fig:gaussian_potential}
\end{figure}

Figure~\ref{fig:gaussian_potential} shows both the Gaussian potential $V(\chi)$ and the corresponding
effective potential $V_{\rm eff}(\chi)$ for representative parameter values.

\paragraph{Background cosmological evolution.}

Defining the density parameters
\begin{equation}
\Omega_i \equiv \frac{\kappa^2\rho_i}{3H^2},
\qquad i=\{\mathrm{r},\mathrm{m},\chi\},
\end{equation}
which satisfy $\Omega_{\rm r}+\Omega_{\rm m}+\Omega_\chi=1$, the Gaussian three-form model reproduces the standard sequence of radiation domination followed by matter domination before entering a late-time dark-energy-dominated phase.

From a dynamical-system perspective \cite{Morais:2016bev,Bouhmadi-Lopez:2016dzw,Bouhmadi-Lopez:2025lzm}, the radiation era corresponds to an early-time repulsive fixed point, followed by matter-dominated saddle configurations. 
Importantly, the small-field and large-field branches are associated with distinct matter saddle points of the autonomous system. 

The small-field branch corresponds to trajectories that remain close to the maximum of the Gaussian potential at $\chi=0$. In this regime, the matter saddle point is located at the local maximum of both $V(\chi)$ and the effective potential $V_{\rm eff}(\chi)$. Since this configuration sits at an unstable extremum, retaining the field in its vicinity for an extended period requires significant fine tuning of the initial conditions. Consequently, although this branch closely mimics $\Lambda$CDM during radiation and matter domination, it is dynamically less generic.

In contrast, the large-field branch corresponds to matter saddle points for which the field amplitude lies far from the maximum of the potential. In this case the field evolves in regions where the effective force does not vanish trivially, leading to a more robust realisation of the matter-dominated era without the need for delicate tuning. From a dynamical viewpoint, this regime appears more natural within the Gaussian three-form framework. A detailed phase-space analysis of the autonomous system, including the explicit characterisation of the radiation and matter fixed points and the distinction between the small- and large-field branches, can be found in Ref.~\cite{Bouhmadi-Lopez:2025lzm}, to which we refer the reader for a comprehensive discussion.

\begin{figure*}
\centering

\begin{subfigure}[t]{0.49\textwidth}
    \centering
    \includegraphics[width=\linewidth]{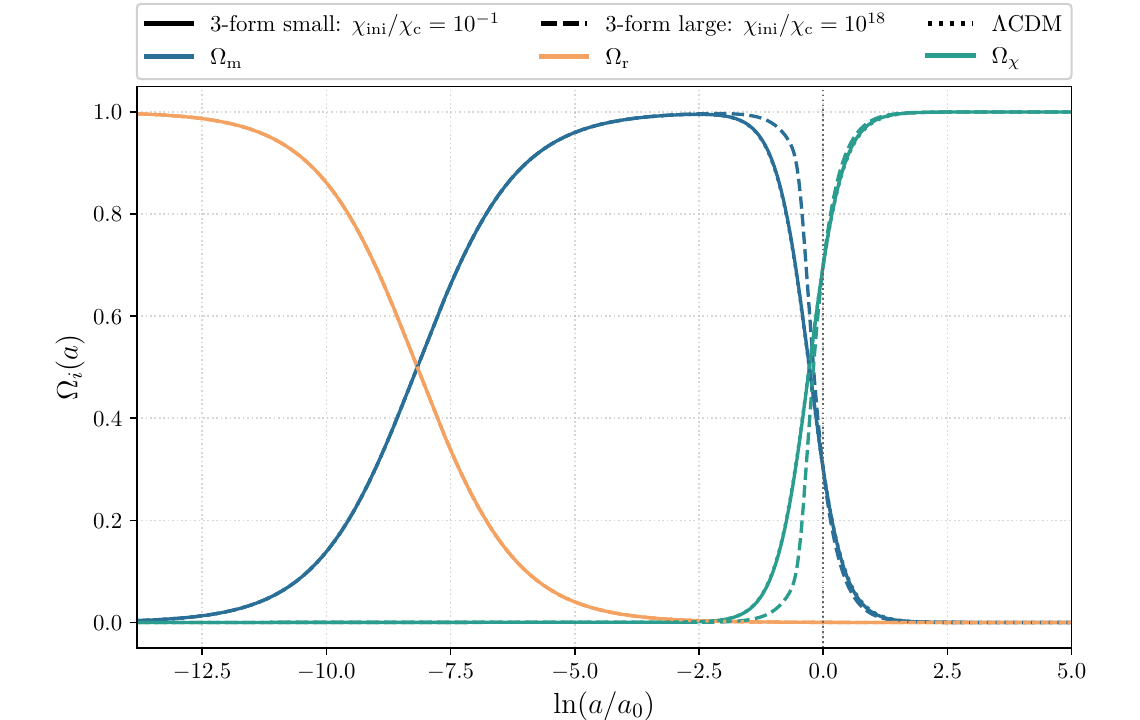}
    \caption{\justifying{Evolution of the density parameters
    $\Omega_{\rm r}$, $\Omega_{\rm m}$ and $\Omega_\chi$.
    We show representative trajectories for the small-field and large-field branches,
    together with the $\Lambda$CDM reference.}}
    \label{fig:omegas_evolution}
\end{subfigure}
\hfill
\begin{subfigure}[t]{0.49\textwidth}
    \centering
    \includegraphics[width=\linewidth]{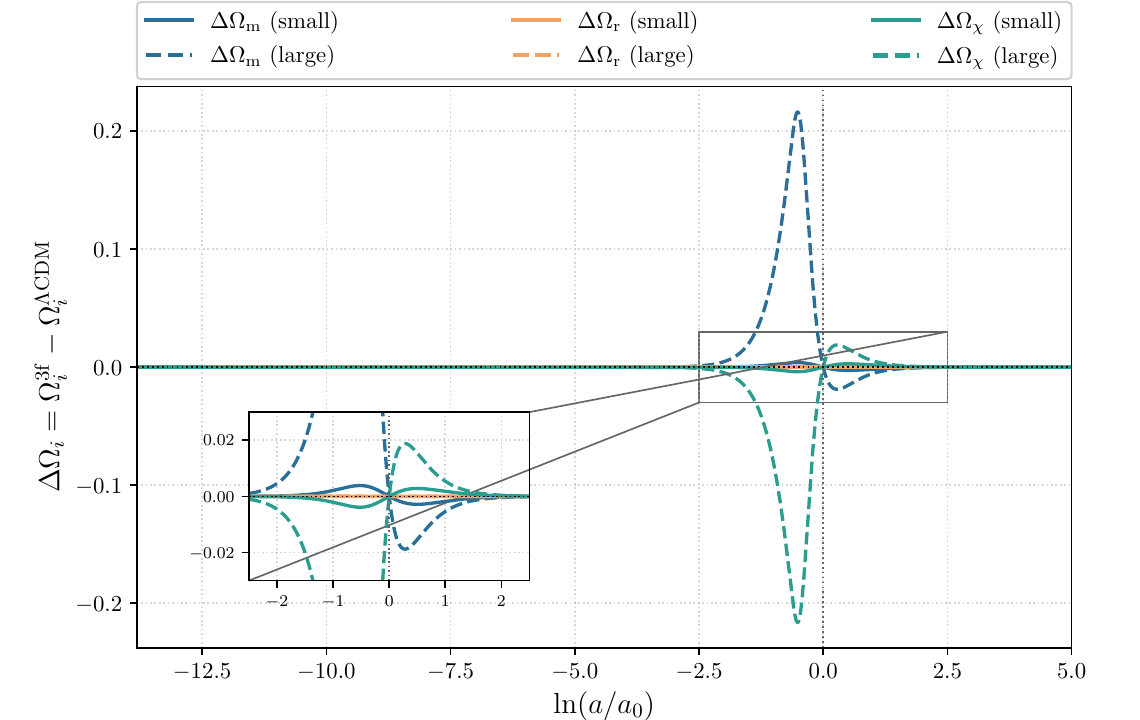}
    \caption{\justifying{Residuals with respect to $\Lambda$CDM:
    $\Delta\Omega_i\equiv \Omega_i^{\rm 3f}-\Omega_i^{\Lambda{\rm CDM}}$
    for $i=\{{\rm r},{\rm m},\chi\}$ (zoomed around late times).}}
    \label{fig:omegas_residuals}
\end{subfigure}

\vspace{0.35cm}

\begin{subfigure}[t]{0.80\textwidth}
    \centering
    \includegraphics[width=\linewidth]{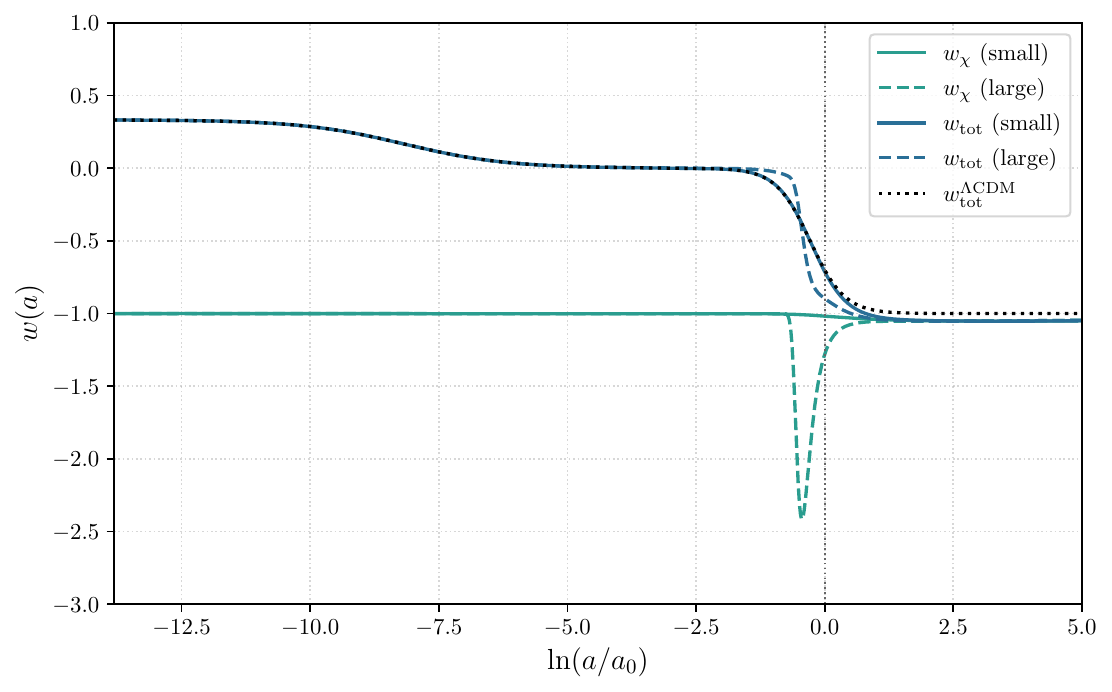}
    \caption{\justifying{Equation of state of the three-form, $w_\chi$,
    and total effective equation of state, $w_{\rm tot}$, for the same representative
    trajectories as in panels (a)-(b), compared with $w_{\rm tot}^{\Lambda{\rm CDM}}$.}}
    \label{fig:eos_wx_wtot}
\end{subfigure}

\caption{\justifying{Representative background evolution in the Gaussian three-form dark-energy model.
Panel (a) shows the density parameters for the radiation, matter and three-form components,
highlighting the existence of small-field and large-field branches.
Panel (b) quantifies their departure from $\Lambda$CDM through $\Delta\Omega_i$.
Panel (c) shows the corresponding evolution of $w_\chi$ and of the total equation of state $w_{\rm tot}$.
}}
\label{fig:background_summary}
\end{figure*}

The background evolution for representative parameter values is shown in figure~\ref{fig:background_summary}. 

Panel (a) displays the density parameters $\Omega_{\rm r}$, $\Omega_{\rm m}$ and $\Omega_\chi$ for both branches, together with the $\Lambda$CDM reference solution. 
The small-field branch closely tracks $\Lambda$CDM throughout the radiation and matter-dominated eras and remains practically indistinguishable from it up to very recent times. 
Only near the present epoch do small deviations emerge, driven by the phantom-like behaviour of the three-form field ($w_\chi<-1$ once the field departs from $\chi=0$).

This behaviour is made more transparent in panel (b), where we plot the residuals 
$\Delta\Omega_i\equiv \Omega_i^{\rm 3f}-\Omega_i^{\Lambda{\rm CDM}}$.
For the small-field branch, the residuals remain negligible during the entire matter-dominated era and grow only at low redshift, reflecting the late onset of phantom dynamics.
In contrast, the large-field branch exhibits appreciable deviations already during the matter epoch, signalling a stronger backreaction of the three-form on the expansion history at intermediate redshifts.

Finally, panel (c) shows the evolution of the three-form equation of state $w_\chi$ together with the total effective equation of state $w_{\rm tot}\equiv p_{\rm tot}/\rho_{\rm tot}$. 
Both branches reproduce the expected sequence $w_{\rm tot}\simeq 1/3$ (radiation) and $w_{\rm tot}\simeq 0$ (matter), followed by a transition to negative values at late times.
However, while the small-field branch remains very close to the $\Lambda$CDM behaviour until recent epochs, the large-field branch enters the phantom regime earlier, which explains the enhanced deviations observed in panel (b).

For the numerical integration shown in figure~\ref{fig:background_summary}, the amplitude $V_\ast$ is fixed so as to satisfy the closure relation $\Omega_{\rm r}+\Omega_{\rm m}+\Omega_\chi=1$ at the present time. 
In practice, small numerical inaccuracies can accumulate during the integration of the background equations and spoil the closure condition if $V_\ast$ is not adjusted consistently. 
We therefore determine $V_\ast$ iteratively so that the present-day density parameters satisfy the flatness constraint to high precision.






\section{Data}\label{observations}
We focus on studying the constraints on the model from an array of cosmological data that represents both high and low redshift and incorporates the discrepancy of the Hubble tension. These data are: Cosmic Microwave Background shift parameters (CMB), baryon acoustic oscillations from the DESI DR2 release (DESIBAO2), Pantheon+ supernovae data, either calibrated with SH0ES distance anchors (PPS) or not (Pan), cosmic chronometers (CC) and Gamma-ray bursts (GRB).

We separate these datasets into 3 groups: the early time, high-$z$ dataset of CMB and DESIBAO2, the high-$H_0$ distance ladder measurement of PPS, and the low-$H_0$ data of CC, GRB, and Pan. Given the direct conflict between the latter two groups, we consider in total 6 combinations of these datasets for tension analysis: DESIBAO2, CMB+DESIBAO2, PPS, CC+GRB+Pan, CMB+DESIBAO2 + PPS and CMB+DESIBAO2+CC+GRB+Pan. We do not perform a fit to only CMB shift parameters as the number of data points (3) is smaller than the number of free parameters of the model (7).

\subsection{CMB shift parameters}

These conform a compression of the high-redshift information of the CMB to three data points that are notably model-independent, especially when changing dark energy models \cite{Wang_2006}. They can be derived from the full information of the CMB. In this work, we employ the distance priors derived in \cite{Zhai_2019}.

\subsection{DESI DR2 BAO data}

Baryon acoustic oscillations are fluctuations in the density distribution of baryon matter at large scales, caused by the propagation of waves in the matter-photon fluid before recombination. They provide a characteristic scale that can be used as a standard ruler, namely the comoving sound horizon at the high-redshift baryon drag epoch, $r_d$.

We use the full BAO Data as reported by the latest data release of the DESI collaboration \cite{Abdul_Karim_2025i,Abdul_Karim_2025ii}. This includes 13 measurements up to $z=2.33$.

\subsection{SNe data}

We use the Pantheon+ Type Ia SNe catalogue, either calibrated with SH0ES Cepheid data (abbreviated as ``PPS''), or without any calibration (abbreviated as ``Pan'') \cite{Pan+S}. It includes a collection with 1701 light curves of SNIa from $z=0.001$ to $z=2.26$. These SNe serve as standard candles due to the relation between their peak brightness and light curve. When the SH0ES calibration is not used (Pan), the supernova sample is effectively calibrated through its combination with cosmic chronometer (CC) data in the joint analysis. 

\subsection{Cosmic chronometers}
\label{ccdata}

We include direct measurements of the Hubble expansion rate obtained with the cosmic chronometer (CC) technique, which relies on the differential ageing of passively evolving galaxies \cite{Jimenez:2001gg}. In this work we adopt the compilation of Ref.~\cite{Moresco:2024wmr}, consisting of $N_{\rm CC}=33$ measurements spanning approximately $0.07 \lesssim z \lesssim 2.0$.

\subsection{Gamma-ray bursts}
\label{grbdata}

To extend background distance information to redshifts well beyond those covered by Type Ia supernovae, we also consider gamma-ray bursts (GRB) as high-redshift standardisable candles. In particular, we use the ``Mayflower'' GRB Hubble-diagram sample of Ref.~\cite{Liu:2014vda}, which contains $N_{\rm GRB}=79$ objects in the range $1.44<z<8.1$.

\section{Statistical methodology}\label{sec:methodology}

To obtain the observational constraints for the model, we perform a Bayesian analysis, obtaining the posterior distribution for both the free parameters of the model and useful derived parameters introduced in section~\ref{sec:prior}, from the dataset summarised in the previous section, expressed as the likelihood functions introduced in section~\ref{sec:likelihoods}. To do so, we employ the \texttt{Cobaya} package \cite{Torrado:2020dgo, Lewis:2002ah, Lewis:2013hha,neal2005takingbiggermetropolissteps, 2019ascl.soft10019T} and its implemented Markov chain Monte Carlo (MCMC) Metropolis sampler to approximate the Bayes formula
\begin{align}
P(\theta_M | \mathcal D) = \frac{ P(\mathcal D | \theta_M) P(\theta_M)}{\int P(\mathcal D | \theta_M) P(\theta_M) d\theta_M}  \,,
\end{align}
where $P(\theta_M | \mathcal D)$ is the posterior of the parameters $\theta_M$ for model $M$ given dataset $\mathcal D$, $P(\mathcal D | \theta_M)$ is the likelihood of $\mathcal D$ as a function of $\theta_M$, and $P(\theta_M)$ is the prior density of the parameter space. To represent the fits, we use the GetDist package \cite{GetDist}.

\subsection{Likelihoods}\label{sec:likelihoods}

In this subsection we specify the likelihoods used to compare the observational measurements with the theoretical predictions of the model. Throughout this work, we assume Gaussian likelihoods for all datasets. The likelihood is expressed in terms of the chi-squared statistic as
\begin{equation}
P(\mathcal D | \theta_M) \propto \exp\!\left(-\frac{\chi^2}{2}\right),
\end{equation}
where $\chi^2$ quantifies the weighted squared differences between the data and the corresponding model predictions given specific parameters $\theta_M$.\footnote{In this work $n$-$\sigma$ refers to the probability according to the chi-square distribution with $n$ being the value of $\chi$.}

\subsubsection{CMB shift parameters}

We employ the distance priors derived in \cite{Zhai_2019}, given by the parameters $\omega_{\mathrm{b0}}=\Omega_{\mathrm{b0}}h^2$,

\begin{eqnarray}\label{Rcmb}
    R(z_*)=\sqrt{\Omega_{\rm m0}H_0^2}(1+z_*)D_A(z_*)/c,
\end{eqnarray}

\begin{eqnarray}\label{lacmb}
    \ell_a(z_*)=(1+z_*)\pi \frac{D_A(z_*)}{r_s(z_*)},
\end{eqnarray}
and their covariance matrix $C_{\mathrm{CMB}}$. Here, $c$ is the speed of light, $D_A(z)$ is the angular diameter distance

\begin{eqnarray}
    D_A(z)=\frac{c}{(1+z)}\int_0^{z}\frac{dz'}{H(z')}
\end{eqnarray}
and $r_s(z)$ is the comoving sound horizon at redshift $z$:

\begin{eqnarray}
    r_s(z)=\int_z^\infty\frac{c}{H(z')}\frac{dz'}{\sqrt{3\left(1+R_d\left(\frac{10^3}{1+z'}\right)\right)}},
\end{eqnarray}
where $R_d$ is given by \cite{Eisenstein:1997ik}

\begin{eqnarray}
    R_d=31.5\omega_{\mathrm{b}}\left(\frac{T_{\mathrm{CMB}}}{2.7\ \mathrm{K}}\right)^{-4}.
\end{eqnarray}
Finally, $z_*$ is the redshift of photon decoupling, which we obtain from the fitting formula\cite{Hu:1995en}:

\begin{eqnarray}
    z_*=1048(1+0.00124\omega_{\mathrm{b}}^{-0.738})(1+g_1\omega_{\mathrm{m}}^{g_2})
\end{eqnarray}

\begin{eqnarray}
    g_1=0.0783\omega_{\mathrm{b}}^{-0.238}(1+39.5\omega_{\mathrm{b}}^{0.763})^{-1}
\end{eqnarray}

\begin{eqnarray}
    g_2=0.560(1+21.1\omega_{\mathrm{b}}^{1.81})^{-1}.
\end{eqnarray}

The final $\chi^2_{\mathrm{CMB}}$ is obtained through

\begin{eqnarray}
    \chi^2_{\mathrm{CMB}}=\Delta X_{\mathrm{CMB}}^{\mathrm{T}}C_{\mathrm{CMB}}^{-1}\Delta X_{\mathrm{CMB}},
\end{eqnarray}
where $X_{\mathrm{CMB}}$ is the difference between the observed values of the parameters $R$, $\ell_a$ and $\omega_{\mathrm{b}}$ and their predictions by the model.

\subsubsection{DESI DR2 BAO data}

We use the full BAO dataset reported in the latest data release of the DESI collaboration \cite{Abdul_Karim_2025i,Abdul_Karim_2025ii}, as implemented in \texttt{Cobaya}. This dataset comprises 13 measurements extending up to $z=2.33$. The observables are provided as combinations of cosmological distances and the sound horizon $r_d$: $D_M(z)/r_d$, $D_H(z)/r_d$, and $D_V(z)/r_d$, where $r_d$ denotes the comoving sound horizon at the baryon drag epoch.\footnote{Here $D_H(z)$ is the Hubble distance at redshift $z$ and is related to the radial BAO measurement, while $D_M(z)$ is the comoving angular diameter distance corresponding to the transverse measurement. The volume-averaged distance $D_V(z)$ combines the previous two and is only used for the lowest-redshift data point.} For completeness, we now define these distances:


\begin{eqnarray}
     D_H(z)=\frac{c}{H(z)},
\end{eqnarray}

\begin{eqnarray}
     D_M(z)=(1+z)D_A(z),
\end{eqnarray}

\begin{eqnarray}
     D_V(z)=(zD_M(z)^2D_H(z))^{1/3}.
\end{eqnarray}

To compute the sound horizon at the drag epoch, $r_d \equiv r_s(z_d)$, we employ the fitting formula presented in \cite{Brieden_2023}:

\begin{eqnarray}
    r_d=147.05\ \mathrm{Mpc}
    \left(\frac{\omega_{\mathrm{b}}}{0.02236}\right)^{-0.13}
    \left(\frac{\omega_{\mathrm{m}}}{0.1432}\right)^{-0.23},
\end{eqnarray}

where we have omitted the factor depending on the effective number of neutrino species, $N_{\rm eff}$, as it is fixed to 3.04 and therefore not varied in the fit\footnote{The full fitting formula for the comoving sound horizon at the baryon drag epoch, $r_d \equiv r_s(z_d)$, reads \cite{Brieden_2023}:
$$r_d=147.05\ \mathrm{Mpc} \left(\frac{\omega_{\mathrm{b}}}{0.02236}\right)^{-0.13} \left(\frac{\omega_{\mathrm{m}}}{0.1432}\right)^{-0.23} \left(\frac{N_{\rm eff}}{3.04}\right)^{-0.1}.$$}.

The total $\chi^2_{\mathrm{BAO}}$ is obtained in an analogous way to that of the CMB:
\begin{eqnarray}
    \chi^2_{\mathrm{BAO}}=
    \Delta X_{\mathrm{BAO}}^{\mathrm{T}}
    C_{\mathrm{BAO}}^{-1}
    \Delta X_{\mathrm{BAO}},
\end{eqnarray}
where $C_{\mathrm{BAO}}$ is the covariance matrix and $\Delta X_{\mathrm{BAO}}$ is the vector of differences between the observed values of the distance ratios and those predicted by the theoretical model.

\subsubsection{SNe data}

We use the Pantheon+ Type Ia SNe likelihood as implemented in \texttt{Cobaya}. The data are effectively given by sets of values for the observed apparent magnitude, heliocentric redshift and CMB frame redshift ($m_i,z_{{\rm hel},i},z_{{\rm CMB},i}$). The observational distance modulus is defined as
\begin{equation}
\mu_{\rm SN}^{\rm obs}=m-M,
\end{equation}
where $M$ is the standardised absolute magnitude.

One can obtain the theoretical distance modulus from the luminosity distance as

\begin{equation}
\mu_{\rm SN}^{\rm th}(z_{\rm hel},z_{\rm CMB}) = 5\log_{10}\!\left[D_L(z_{\rm hel},z_{\rm CMB})\right] + \mu_0,
\end{equation}
where $\mu_0 \equiv 5\log_{10}\!\left[cH_0^{-1}\,{\rm Mpc}^{-1}\right]+25$ collects the unit conversion from Mpc to 10pc and 

\begin{equation}
D_L(z_{\rm hel},z_{\rm CMB})=(1+z_{\rm hel})\int_0^{z_{\rm CMB}} \mathrm{d}z'\,\frac{H_0}{H(z')}
\end{equation}
is the dimensionless Hubble free luminosity distance. 

When the Pantheon+ \& SH0ES compilation is adopted in the analysis (which is not the case in all our runs, since in some cases we instead combine Pantheon+ with Cosmic Chronometers), the dataset includes a set of Cepheid-host (calibrator) supernovae from the SH0ES distance ladder. For Hubble-flow supernovae, the standard cosmological prediction above is employed. For Cepheid-host supernovae, however, the distance entering the likelihood is replaced by the measured Cepheid-based host distance. This calibration is consistently incorporated through the full Pantheon+ \& SH0ES covariance matrix, which accounts for statistical and systematic uncertainties as well as correlations among supernovae.

We can then define the residual vector $\Delta\mu_{\rm SN}=\mu_{\rm SN}^{\rm obs}-\mu_{\rm SN}^{\rm th}$ and the Pantheon+ \& SH0ES $\chi^2$ as
\begin{equation}
\chi^2_{\rm PPS}=\Delta\mu_{\rm SN}^{\rm T}\,C_{\rm PPS}^{-1}\,\Delta\mu_{\rm SN}  \,,
\end{equation}
where $C_{\rm PPS}^{-1}$ is the covariance matrix of the Pantheon+ \& SH0ES dataset.

On the other hand, if we remove the SNe anchors, there are no longer any observables from the SNe-hosting galaxies that can determine the standardised absolute magnitude of type Ia SNe. $M$ is therefore a flat direction, i.e., $\Delta\chi^2 = 0$ when varying $M$. In practice, this is accounted for by removing the corresponding constant mode from the inverse covariance matrix, leading to
\begin{align}
\chi^2_{\rm Pan} &= \Delta\mu_{\rm SN}^{\rm T} \, C_{\rm Pan}^{-1} \, \Delta\mu_{\rm SN}  \,,\\
C_{\rm Pan}^{-1} &= C_{\rm PPS}^{-1} - \hat 1^{\rm T}  C_{\rm PPS}^{-1} \hat 1  \,,
\end{align}
where $\hat 1 = N_{\rm SN}^{-1/2} ( 1, 1,\dots,1)$ is the normalised $1$-vector with $N_{\rm SN}$ the size of the Pantheon+ catalogue.

\subsubsection{Cosmic chronometers}

For the cosmic chronometer data, the expansion rate is inferred from the relation

\begin{equation}
H(z) = -\frac{1}{1+z}\,\frac{\mathrm{d}z}{\mathrm{d}t},
\end{equation}
providing direct measurements of $H(z)$.

We adopt the compilation of Ref.~\cite{Moresco:2024wmr}, consisting of $N_{\rm CC}=33$ measurements spanning approximately $0.07 \lesssim z \lesssim 2.0$. Each data point is given as $(z_i, H_i, \sigma_{H_i})$, where $H_i$ is the observed expansion rate at redshift $z_i$ and $\sigma_{H_i}$ its reported uncertainty. The theoretical prediction is denoted $H_{\rm th}(z_i)$ and is obtained from the background evolution of the model.

Assuming the measurements to be uncorrelated, the corresponding chi-squared statistic is

\begin{equation}
\chi^2_{\rm CC}=\sum_{i=1}^{N_{\rm CC}}
\frac{\left[H_i - H_{\rm th}(z_i)\right]^2}{\sigma_{H_i}^2}.
\end{equation}

\subsubsection{Gamma-ray bursts}

The GRB dataset is provided in terms of the observed distance modulus $\mu_{\rm GRB}^{\rm obs}$ at a certain redshift and its associated uncertainties. For a given cosmological model, we compute the theoretical prediction $\mu_{\rm GRB}^{\rm th}(z)$ from the luminosity distance
in an analogous way to the one used for SNe, except that there is no need to convert from $z_{\rm hel}$ to $z_{\rm CMB}$ in the definition of $D_L$.

Defining the residual vector 
\[
\Delta\mu_{\rm GRB}=\mu_{\rm GRB}^{\rm obs}-\mu_{\rm GRB}^{\rm th},
\]
the corresponding chi-squared statistic is

\begin{equation}
\chi^2_{\rm GRB}=\Delta\mu_{\rm GRB}^{\rm T}\,C_{\rm GRB}^{-1}\,\Delta\mu_{\rm GRB},
\end{equation}
where $C_{\rm GRB}$ is the covariance matrix of the GRB sample. In this work, we assume it to be diagonal, $C_{\rm GRB}=\mathrm{diag}(\sigma_{\mu,i}^2)$, with $\sigma_{\mu,i}$ the reported uncertainties of each data point.

Since $\mu_{\rm GRB}^{\rm th}$ depends on the nuisance parameter $\mu_0$, we marginalise analytically over $\mu_0$. The resulting marginalised statistic reads \cite{Conley_2010}

\begin{equation}
\widetilde{\chi}^2_{\rm GRB} =
A_{\rm GRB}
+\ln\!\left(\frac{E_{\rm GRB}}{2\pi}\right)
-\frac{B_{\rm GRB}^2}{E_{\rm GRB}},
\end{equation}

where

\begin{equation}
\begin{aligned}
A_{\rm GRB} &= 
\Delta\widetilde{\mu}_{\rm GRB}^{\rm T}
C_{\rm GRB}^{-1}
\Delta\widetilde{\mu}_{\rm GRB}, \\
B_{\rm GRB} &=
\Delta\widetilde{\mu}_{\rm GRB}^{\rm T}
C_{\rm GRB}^{-1}
1, \\
E_{\rm GRB} &=
1^{\rm T}
C_{\rm GRB}^{-1}
1, \\
\Delta\widetilde{\mu}_{\rm GRB} &=
\Delta\mu_{\rm GRB}(\mu_0=0).
\end{aligned}
\end{equation}

\subsection{Prior specification}\label{sec:prior}

The free parameters of the model include the usual present time quantities $H_0=h\cdot100\ \mathrm{km/Mpc}$, $\Omega_{\mathrm{m0}}$ and $\Omega_{\mathrm{b0}}$\footnote{To compute the present radiation density $\Omega_{\mathrm{r0}}$, we first obtain the redshift of equality matter-radiation through $z_{eq}=2.5\cdot10^4\Omega_{\mathrm{m0}}h^2\left(\frac{T_{\mathrm{CMB}}}{T}\right)$ \cite{Eisenstein:1997ik} with $T_{\mathrm{CMB}}=2.7255\ \mathrm{K}$ \cite{Fixsen:2009ug} and then fix it with $\Omega_{\mathrm{r0}}=\Omega_{\mathrm{m0}}/(1+z_{eq})$. The dark energy density is obtained through the constraint of a flat Universe: $\Omega_{\mathrm{DE,0}}=1-\Omega_{\mathrm{m0}}-\Omega_{\mathrm{r0}}$}, as well as the parameters of the three-form: $\log_{10} \xi$, $\log_{10}(a_i^3\sqrt{\xi}\kappa\chi_i)$, $\mathrm{sign}(\Pi_i)\mathrm{KE}_i/\rho_{\rm DE,0}$, and $(\textrm{KE}_i+V_*)/\rho_{DE,0}$, where $\Pi_i=(\dot{\chi_i}+3H_i\chi_i)$ is the momentum\footnote{The momentum defined here is canonical to $\phi = a^3 \chi$. For more details on the definition of three-form momentum please see \cite{Bouhmadi-Lopez:2018lly}.}, $\textrm{KE}_i=\Pi_i^2/2$  is the kinetic energy of the three-form field at the initial time and $\rho_{\rm DE,0}$ its current energy density. These parameters are chosen to determine the shape of the potential by fixing $\xi$ and $V_*$ and also the initial conditions of the field $\chi$ and its derivative $\dot{\chi}$ at an early time $a_i=10^{-6}$.  In comparison with \cite{Bouhmadi-Lopez:2025lzm}, we switch from parametrising the potential height $V_*$ to $(\textrm{KE}_i+V_*)/\rho_{DE,0}$ for faster convergence, as its distribution highly concentrates around unity \cite{Bouhmadi-Lopez:2025lzm}. This faster convergence allows us to reach the stopping criteria for all dataset combinations.

The parameters and their priors are summarised in table \ref{tab:param priors}. In addition, we impose the cut $40<H_0<100$, as well as a Big Bang Nucleosynthesis (BBN) prior on the baryonic density $\omega_{\rm b0}=\Omega_{\rm b0} h^2=0.02218\pm0.00055$ \cite{Schoneberg:2024ifp}. 

\begin{table}[t]
\centering
\begin{tabular}{c c c}
\hline\hline
Parameter & Min. & Max. \\
\hline\hline
$H_0$  & $40$ & $100$ \\
$\Omega_{m0}$  & $0.1$ & $0.9$ \\
$\Omega_{b0}$  & $0.01$ & $0.09$ \\
\hline
$\log_{10}(\xi)$ & -7 & $\log_{10}(9/2)$\footnote{The prior chosen here is based on the analysis in Ref.~\cite{Bouhmadi-Lopez:2025lzm}.} \\
$\log_{10}(a_i^3\sqrt{\xi}\kappa\chi_i)$ & -3.5 & 0.5 \\
$\mathrm{sign}(\Pi_i)\mathrm{KE}_i/\rho_{\rm DE,0}$ & -1 & 1 \\
$(\textrm{KE}_i+V_*)/\rho_{DE,0}$ & 0.5 & 1.5 \\
\hline\hline
\end{tabular}
\caption{\justifying{The parameters of $\Lambda$CDM and those of the three-form used in the analysis, as well as the minimum and maximum of their flat priors.}}
\label{tab:param priors}
\end{table}

We set the initial conditions at $a_i$ with these parameters and solve the equation of motion towards the present, obtaining a cosmological evolution for the model. We also need to ensure that the current Hubble parameter obtained through this process, $H_{0, \mathrm{EOM}}$, coincides with the free parameter $H_0$, that appears only as a scale. To do so, we include a Gaussian likelihood on $\varepsilon_{H_0}=1-\frac{H_{0,\mathrm{EOM}}^2}{H_0^2}$ of width $10^{-3}/\sqrt{2}$. We disregard the likelihood of this constraint when calculating the relevant quantities, effectively treating it as a prior on $\varepsilon_{H_0}$. 

\subsection{Model-comparison criteria}\label{sec:model-comparison}

We compare the Gaussian three-form model against $\Lambda$CDM using appropriate statistical diagnostics, including information criteria and Bayesian evidence. The specific criteria adopted are the corrected Akaike information criterion ($\mathrm{AIC_C}$), the Bayesian information criterion (BIC) for the frequentist approach, and the Bayes evidence ($\mathrm{B}$), the widely applicable information criterion ($\mathrm{WAIC}$), the deviance information criterion ($\mathrm{DIC}$) for the Bayesian approach.\footnote{There is an extra factor of 2 for the IC values presented in this work in comparison with \cite{Chiang:2025qxg,Bouhmadi-Lopez:2025lzm}.\label{foot:2factor}} These criteria include penalty terms for the number of parameters to 

counteract the advantage a model may possess over models with less parameters. We present multiple criteria to better assess the trade-off between three-form dark energy model and $\Lambda$CDM.

The frequentist approach focuses on the expansion of the likelihood function around the maximum. We utilise the minimiser of \texttt{Cobaya} to identify the maximum of the likelihood\footnote{The Gaussian prior on $\varepsilon_{H_0}$ is considered part of the likelihood during the minimisation to maintain the constraint.}, the so-called maximum likelihood estimate ($\rm mle$), and to extract the following criteria.

The $\mathrm{AIC_C}$ is obtained by minimising the difference between the true data distribution and the model distribution and is defined as \cite{Liddle_2007,Kenneth:2004a}
\begin{equation}
    \mathrm{AIC_C}=-2\ln P(D|\theta_{M,\,{\rm mle}}) + 2k + \frac{2k^2+2k}{n-k-1},
\end{equation}
where $P(D|\theta_{M,\,{\rm mle}})$ is the maximum value of the likelihood reported by the minimiser, $n$ is the total number of data points included and $k$ is the number of free parameters in the model: 3 for background $\Lambda$CDM and 6 for the three-form model.\footnote{While there are 4 more prior parameters for the three-form model than background $\Lambda$CDM, one of them is fully constrained by the current total energy density condition $\varepsilon_{H_0} = 0$. This is reflected by the negligible contribution of the Gaussian likelihood on $\varepsilon_{H_0}$ at the maximum of the posterior. }
For combinations of datasets with a high number of data points ($n\gg k^2$), the last term can be dropped, recovering the uncorrected $\mathrm{AIC}=-2\ln P(D|\theta_{M,\,{\rm mle}}) + 2k$.

The BIC derives from approximating the evidence ratios and is defined as \cite{Liddle_2007,Kenneth:2004b}
\begin{equation}
    \mathrm{BIC}=-2\ln P(D|\theta_{M,\,{\rm mle}}) + k\ln n.
\end{equation}
It favours simpler models over models with more parameters more strongly than $\rm AIC_C$, especially for large number of data points.

On the other hand, the Bayesian approach targets the posterior, leading to the following criteria:
\begin{equation}
\begin{aligned}
-\ln{\rm B} (D|M) &\equiv - 2 \ln \int P(D|\theta_M) P(\theta_M) d\theta_M  \label{eq: -lnB}\\
&= -2 \ln V_M + 2 \ln \left< \left( P(D|\theta_M) \right)^{-1} \right>_{M|D}  \,, \\
{\rm DIC} (D|M) &\equiv 2 \ln P(D|\theta_{M,\,{\rm map}}) - 4 F(D|M)  \,,\\
{\rm WAIC} (D|M) &\equiv - 2 F(D|M) + {\rm BMD} (D|M)  \,,\\
F(D|M) &\equiv \left< \ln P(D|\theta_M) \right>_{M|D}  \,,\\
{\rm BMD} (D|M) &\equiv 2 \left< \left( \ln P(D|\theta_M) \right)^2 \right>_{M|D}  \\
&- 2\left< \ln P(D|\theta_M) \right>_{M|D}^2  \,,  
\end{aligned}
\end{equation}
Here $\left< \ldots \right>_{M|D} \equiv \int (\ldots) P(\theta_M|D) d\theta_M$ is the MCMC mean, $V_M \equiv \int P(\theta_M) d\theta_M$ is the prior volume\footnote{We correct an erroneous sign error for the prior volume term in \cite{Chiang:2025qxg,Bouhmadi-Lopez:2025lzm}. This would not affect the conclusion in these works though as the prior volume term is negligible in these models.}, $F$ is the deviance, BMD stands for the Bayesian model dimension, and $\rm map$ is the abbreviation of maximum-a-posteriori, i.e., the maximum of the posterior distribution.

For all of these criteria, a higher $\rm IC$ value indicates that observations disfavour the model. We define $\Delta \mathrm{IC}=\mathrm{IC}_{\mathrm{three-form}}-\mathrm{IC}_{\mathrm{\Lambda CDM}}$. Therefore, for each of these quantities, a positive (negative) value indicates preference for $\Lambda$CDM (three-form model).
Our choice of normalisation is such that all these probes function identically as the Bayesian ratio test, i.e., the Jeffreys scale as shown in table~\ref{tab: Jeffrey’s scale}.

\begin{table}[]
\centering
\begin{tabular}{cr}
\hline
\hline
$\Delta \ln{\rm B}$    & Interpretation    \\
\hline
$>10$        & Strongly disfavoured / tensioned   \\
$5\sim10$  & Moderately disfavoured / tensioned \\
$2\sim5$  & Weakly disfavoured / tensioned     \\
$-2\sim2$   & Inconclusive                       \\
$-5\sim-2$& Weakly favoured / aligned          \\
$-10\sim-5$& Moderately favoured / aligned      \\
$<-10$       & Strongly favoured / aligned        \\
\hline
\hline
\end{tabular}
\caption{\label{tab: Jeffrey’s scale}\justifying{
Jeffreys' scale for evaluating the evidence of model $M_1$ over $M_2$ or the tension between datasets.}
}
\end{table}

In similar vein, we introduce the tension probes Bayes ratio ($\ln \mathrm R$), goodness of fit ($\rm GoF$) and suspiciousness ($\rm S$) to detect the inherent tension between datasets $D_1$ and $D_2$ from model $M$'s point of view:

\begin{align}
{\rm GoF} (D_1, D_2 | M) &\equiv - 2 \ln P(D_1 D_2 |\theta_{M,\,{\rm map}})  \\*
&+ 2 \ln P(D_1 |\theta_{M,\,{\rm map}}) \nonumber\\*
&+ 2 \ln P(D_2 |\theta_{M,\,{\rm map}})  \,,  \nonumber\\*
{\rm S} (D_1, D_2 | M) &\equiv - 2 F(D_1 D_2 | M)  \\*
&+ 2 F (D_1 | M) + 2 F (D_2 | M)  \,.\nonumber
\end{align}
The value extracted follows a $\rm BMD$-dimensional $\chi^2$ distribution, but one may translate it to the standard $n-\sigma$ value via
\begin{equation}
{\rm CDF}_1^{-1}\!\left[{\rm CDF}_{{\rm BMD}_{\rm eff}}\!\left(\sqrt{{\rm BMD}_{\rm eff}+z}\right)\right],
\end{equation}
where $z$ is the goodness of fit or the suspiciousness,
\begin{align}
{\rm BMD}_{\rm eff}={}&{\rm BMD}(D_1|M)+{\rm BMD}(D_2|M)\nonumber\\
&-{\rm BMD}(D_1D_2|M)
\end{align}
is the effective Bayesian model dimension, and
\begin{equation}
{\rm CDF}_d(z)=\int_0^z e^{-\chi^2}\chi^{2d-2}\,\mathrm{d}\chi^2
\end{equation}
is the cumulative distribution function of a $d$-dimensional Gaussian distribution.

\subsection{Stopping criteria and the merging of chains}

Multiple chains enable us to extract the statistical variance of the derived quantities such as IC values and conduct the variance of mean analysis a la $R-1$. We consider each chain with equal weight, leading to the mean and the standard variance reported in this work:
\begin{align}
\text{mean} (\mathcal O (\Theta)) &\equiv \left< \mathcal O (\Theta) \right> \nonumber\\
&= \frac{1}{N_c} \sum_{i=1}^{N_c} \left< \mathcal O (\Theta) \right>_i  \,,\\
\text{var}\!\left(\mathcal O (\Theta)\right) &= \frac{1}{N_c} \sum_{i=1}^{N_c} \left< \left( \mathcal O (\Theta) - \left< \mathcal O (\Theta) \right>_i \right)^2 \right>_i  \\
&+ \frac{1}{N_c - 1} \sum_{i=1}^{N_c} \left< \left( \left< \mathcal O (\Theta) \right>_i - \left< \mathcal O (\Theta) \right> \right)^2 \right>_i  \,,\nonumber
\end{align}
where the angled bracket with subscript $i$ denotes the mean over the $i$-th chain, $N_c$ is the number of chains. We apply the procedure to Bayesian probes after converting to 1-$d$ Gaussian distribution. For other quantities, we first merge the chains with equal weight and then utilise \texttt{getdist} to generate either the $68\%$ and $95\%$ confidence level bound or the posterior distribution.

We utilise a similar variance analysis for the stopping criteria of Gelman-Rubin\footnote{The quantity $R-1$ is defined as the greatest eigenvalue of $S.C.S^T$
where $C
$ is the covariance of the mean of the parameters, $S$ is a matrix satisfying $S. M . S^T = I$ (e.g. inverse of the Cholesky), and $M$ is the mean of the covariance of the parameters
. Statistical operations for $C$ and $M$ are defined as first operating on each chain after burn-in removal and then over all chains (with equal weight) for inter-chain $R-1$. For intra-chain $R-1$, we first remove 20\% of a chain (without burn-in removal) and break it into 4 segments, then apply the first statistical operation on each segment and the second over all segments. 

}, with $R-1 < 0.01$ both intra-chain and inter-chain.

We remove the burn-in phase of MCMC by discarding the earliest $30\%$ samples per chain.

\section{Results}\label{fits}

In the current section, we present the results of the fits using the data presented above for a three-form dark energy model endowed with a Gaussian potential.

\subsection{Parameter constraints}\label{sec:params}

\begin{figure*}
\centering
\hfill
\begin{subfigure}[t]{0.48\textwidth}
    \centering
    \includegraphics[width=\linewidth]{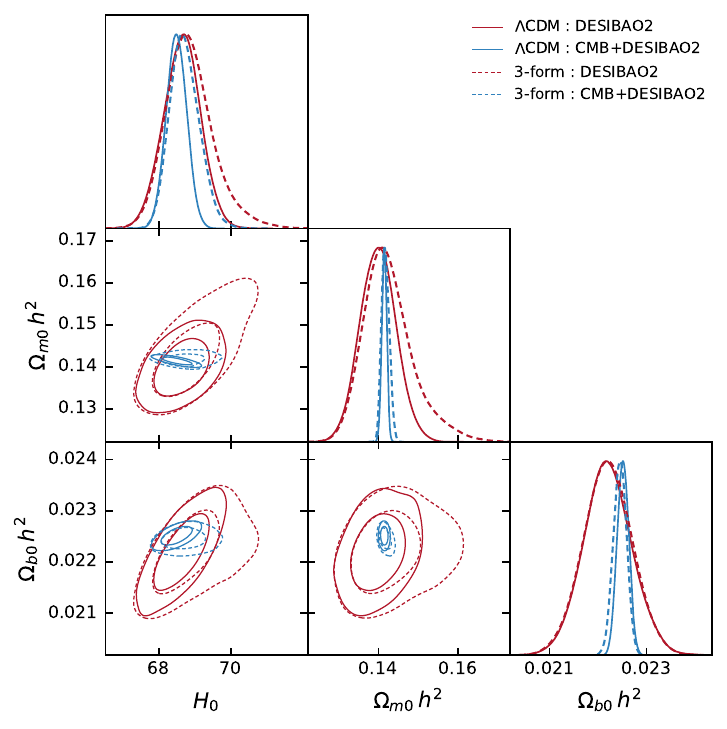}
    \caption{DESIBAO2 and CMB+DESIBAO2.}
    \label{fig:contour_bao}
\end{subfigure}
\hfill\null
\\[6pt]
\begin{subfigure}[t]{0.48\textwidth}
    \centering
    \includegraphics[width=\linewidth]{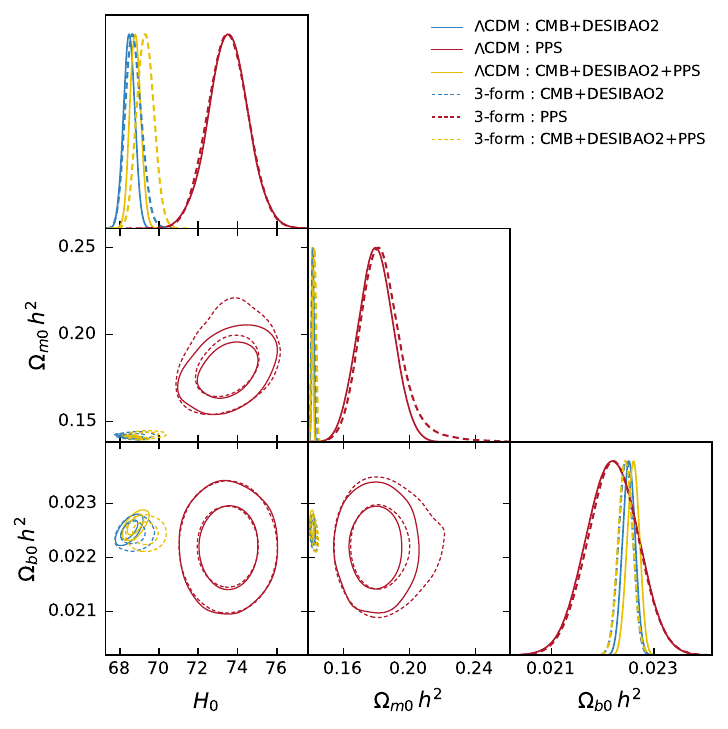}
    \caption{PPS, CMB+DESIBAO2, and CMB+DESIBAO2+PPS.}
    \label{fig:contour_pps}
\end{subfigure}
\hfill
\begin{subfigure}[t]{0.48\textwidth}
    \centering
    \includegraphics[width=\linewidth]{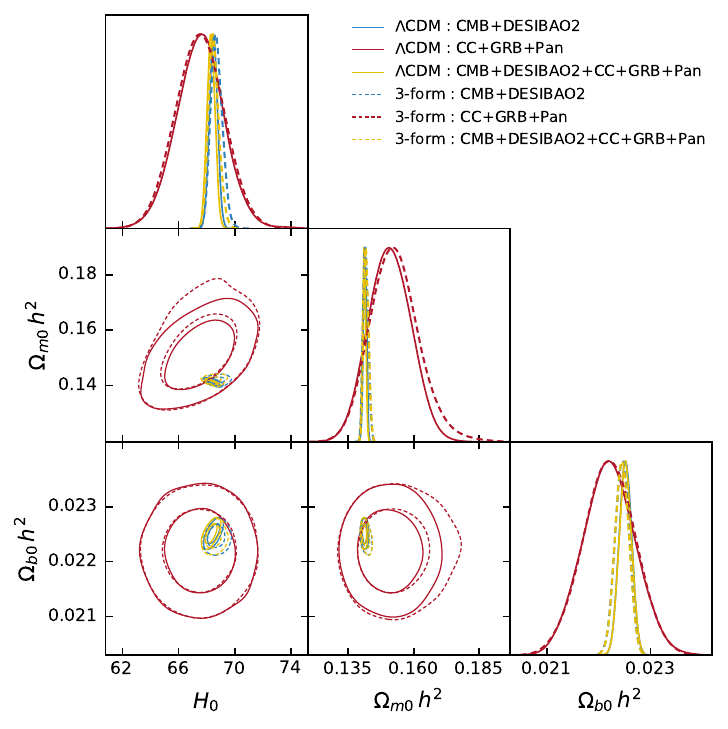}
    \caption{CC+GRB+Pan, CMB+DESIBAO2, and CMB+DESIBAO2+CC+GRB+Pan.}
    \label{fig:contour_cc}
\end{subfigure}
\caption{\justifying{68\% and 95\% posterior distributions for $H_0$, $\Omega_\mathrm{m0}h^2$, and $\Omega_\mathrm{b0}h^2$. Solid lines: $\Lambda$CDM; dashed lines: three-form model. \textit{Top}: DESIBAO2 and CMB+DESIBAO2 (early-time). \textit{Bottom left}: PPS combinations. \textit{Bottom right}: CC+GRB+Pan combinations.}}
\label{fig:contours}
\end{figure*}

The mean and standard deviation on the cosmological parameters and statistical probes are shown in table \ref{tab:LCDM_params} for $\mathrm{\Lambda}$CDM and in table \ref{tab:3form_params} for the three-form model. The posterior distributions of $H_0$, $\Omega_\mathrm{m0}h^2$, and $\Omega_\mathrm{b0}h^2$ are shown for all six dataset combinations and both models in figures~\ref{fig:contour_bao}, \ref{fig:contour_pps}, and~\ref{fig:contour_cc}.

The standard cosmological parameters are consistent between the two models for every dataset combination. It can also be seen that CC+GRB+Pan is consistent with CMB+DESIBAO2 for both models due to its wide posterior. Meanwhile, PPS shows the known $H_0$ tension against CMB+DESIBAO2, which seems to be somewhat alleviated by the three-form model, as it increases the allowed values of $H_0$. Still, the three-form model does not seem able to improve the $\Omega_{m0}$ fit, which does not shift with respect to the $\Lambda$CDM model results.
As expected, the baryon density $\Omega_\mathrm{b0}h^2$ distribution is that of the BBN prior for combinations without CMB shift parameters, and tightened when CMB data are included.

\begin{figure*}
\centering
\begin{subfigure}[t]{0.45\textwidth}
    \centering
    \includegraphics[width=\linewidth]{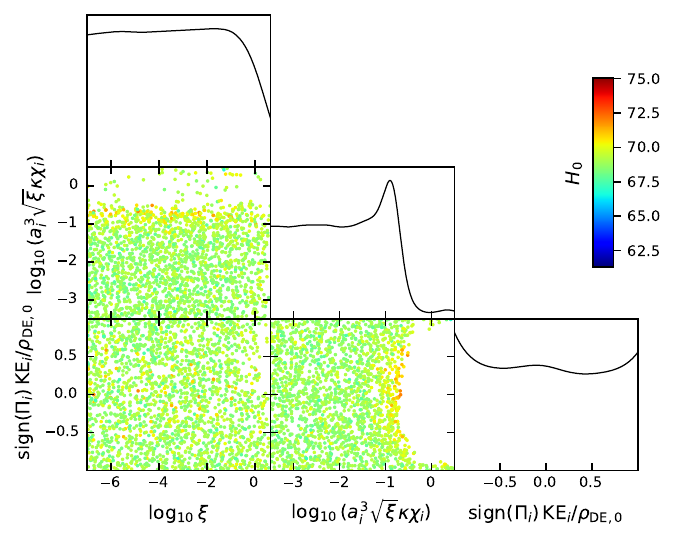}
    \caption{DESIBAO2}
    \label{fig:scatter_bao_only}
\end{subfigure}
\hfill
\begin{subfigure}[t]{0.45\textwidth}
    \centering
    \includegraphics[width=\linewidth]{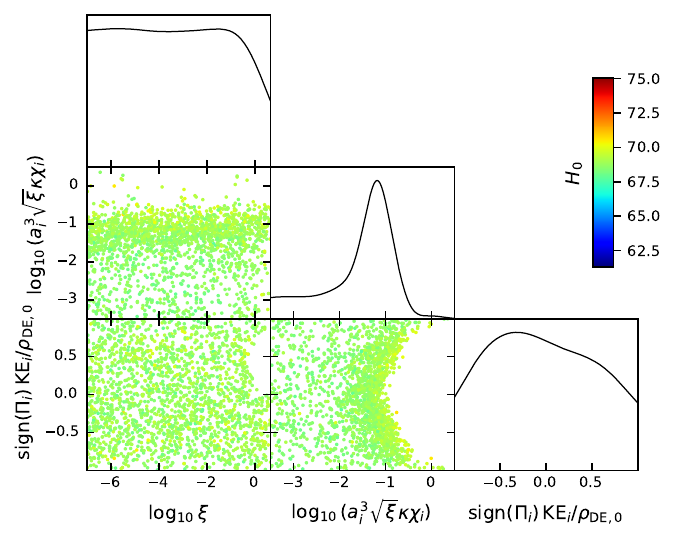}
    \caption{CMB+DESIBAO2}
    \label{fig:scatter_bao}
\end{subfigure}
\\[0pt]
\begin{subfigure}[t]{0.45\textwidth}
    \centering
    \includegraphics[width=\linewidth]{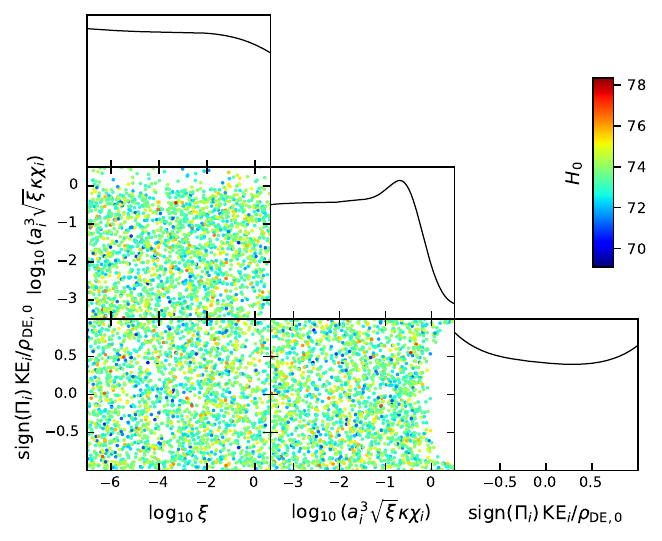}
    \caption{PPS}
    \label{fig:scatter_pps}
\end{subfigure}
\hfill
\begin{subfigure}[t]{0.45\textwidth}
    \centering
    \includegraphics[width=\linewidth]{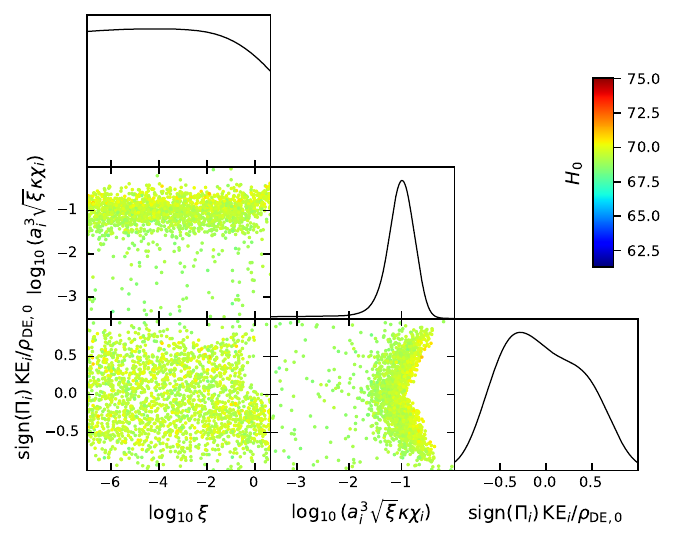}
    \caption{CMB+DESIBAO2+PPS}
    \label{fig:scatter_cpps}
\end{subfigure}
\\[0pt]
\begin{subfigure}[t]{0.45\textwidth}
    \centering
    \includegraphics[width=\linewidth]{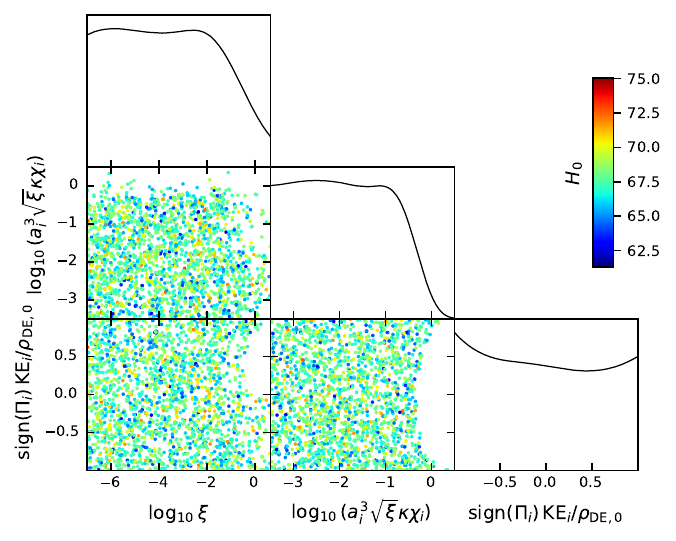}
    \caption{CC+GRB+Pan}
    \label{fig:scatter_cc}
\end{subfigure}
\hfill
\begin{subfigure}[t]{0.45\textwidth}
    \centering
    \includegraphics[width=\linewidth]{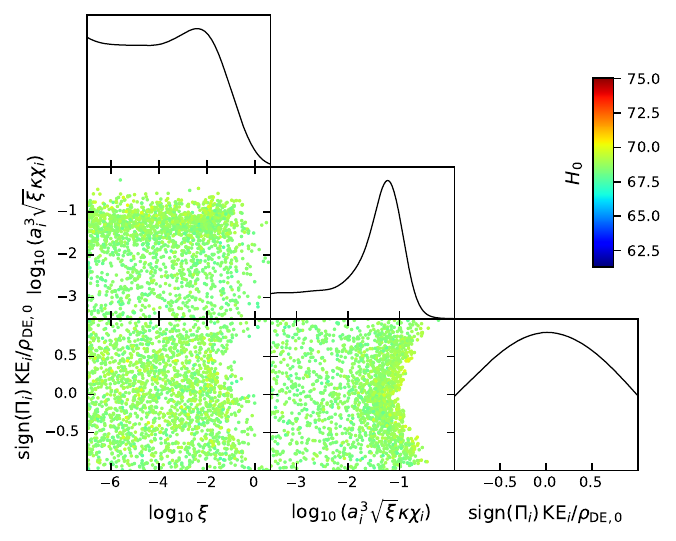}
    \caption{CMB+DESIBAO2+CC+GRB+Pan}
    \label{fig:scatter_ccc}
\end{subfigure}
\caption{\justifying{Posterior scatter plots of the three-form field parameters colour-coded by $H_0$ (note the different colour scale in the case of PPS): potential width $\log_{10}\xi$, initial field strength $\log_{10}(a_i^3\sqrt{\xi}\kappa\chi_i)$, and initial kinetic energy term with the sign of the momentum $\mathrm{sign}(\Pi_i)\mathrm{KE}_i/\rho_{\rm DE,0}$, where the momentum is defined as $\Pi \equiv \dot\chi + 3 H \chi$.
\textit{Left column}: individual datasets (DESIBAO2, PPS, CC+GRB+Pan).
\textit{Right column}: combined datasets with CMB+DESIBAO2.}}
\label{fig:scatter}
\end{figure*}

Turning to the three-form field parameters, their distribution for the six dataset combinations considered is plotted in figures~\ref{fig:scatter_bao_only} -~\ref{fig:scatter_ccc}. In them, it can be seen that $\log_{10}\xi$ is consistent with the lower boundary of its prior ($\xi\to 0$, the cosmological constant limit) for all dataset combinations, with an upper end dominated by the prior or a weak limit depending on the dataset. The $\log_{10}(a_i^3\sqrt{\xi}\kappa\chi_i)$ parameter similarly has upper limits for all datasets, but the late-time only datasets do not impose a lower constraint. On the other hand, the fits that include early-time DESIBAO2 and CMB start showing preference for some central values with a lower tail of varying height. In the case of CMB+DESIBAO2+PPS, the tail is extremely weak and we can confidently say that there is a preference for a central value. The parameter $\mathrm{sign}(\Pi_i)\mathrm{KE}_i/\rho_{\rm DE,0}$ only shows a weak constraint for combinations with CMB data and is unconstrained for the rest, and from the tables we can see that, as expected, the parameter $(\mathrm{KE}_i+V_*)/\rho_{\mathrm{DE},0}$ clusters around $1$ across all fits.

\begin{table*}
\centering
\resizebox{\textwidth}{!}{
\begin{tabular}{lcccccc}
\hline
 & DESIBAO2 & CMB+DESIBAO2 & PPS & CMB+DESIBAO2+PPS & CC+GRB+Pan & CMB+DESIBAO2+CC+GRB+Pan \\
\hline
$H_0$
& $68.63\pm0.51$
& $68.48\pm0.29$
& $73.5\pm1.0$
& $68.80^{+0.26}_{-0.29}$
& $67.5\pm1.7$
& $68.31\pm0.29$ \\

$\Omega_\mathrm{m0}$
& $0.2969\pm0.0085$
& $0.3016\pm0.0036$
& $0.332\pm0.018$
& $0.2978\pm0.0034$
& $0.331\pm0.018$
& $0.3023\pm0.0037$ \\

$\Omega_\mathrm{b0}$
& $0.04712\pm0.00074$
& $0.04800\pm0.00032$
& $0.0411\pm0.0015$
& $0.04775\pm0.00031$
& $0.0488^{+0.0024}_{-0.0028}$
& $0.04825\pm0.00033$ \\

$\Omega_\mathrm{m0} h^2$
& $0.1399\pm 0.0045$ 
& $0.14142\pm0.00062$
& $0.180\pm0.011$
& $0.14096\pm 0.00060$
& $0.1509\pm0.0081$
& $0.14106\pm0.00063$ \\

$\Omega_\mathrm{b0} h^2$
& $0.02220\pm0.00050$
& $0.02252\pm0.00012$
& $0.02219\pm0.00050$
& $0.02260\pm0.00012$
& $0.02220\pm0.00050$
& $0.02251\pm0.00012$ \\
\hline
$\mathrm{AIC_C}$
  & $18.95039\pm0.00027$
  & $23.261\pm0.013$
  & $1458.0311376\pm0.0000026$
  & $1504.1771\pm0.0049$
  & $1457.9615\pm0.0011$
  & $1477.54825\pm0.00042$ \\
BIC
  & $17.97858\pm0.00027$
  & $23.579\pm0.013$
  & $1474.3339098\pm0.0000026$
  & $1520.5081\pm0.0049$
  & $1474.4564\pm0.0011$
  & $1494.06967\pm0.00042$ \\
DIC
  & $14.21\pm0.28$
  & $20.80\pm0.19$
  & $1456.07\pm0.12$
  & $1503.756\pm0.075$
  & $1457.80\pm0.46$
  & $1477.53\pm0.36$ \\
WAIC
  & $14.12\pm0.29$
  & $20.64\pm0.32$
  & $1456.02\pm0.13$
  & $1503.53\pm0.19$
  & $1457.93\pm0.99$
  & $1477.72\pm0.98$ \\
$-\ln{\rm B}$
  & $14.31\pm0.56$
  & $22.1\pm2.4$
  & $1455.96\pm0.10$
  & $1503.28\pm0.51$
  & $1459.2\pm1.7$
  & $1479.2\pm3.5$ \\
\hline
\end{tabular}
}
\caption{\justifying{$68\%$ confidence intervals on cosmological parameters and information criteria for $\Lambda$CDM across all six dataset combinations.}}\label{tab:LCDM_params}

\end{table*}

\begin{table*}
\centering
\resizebox{\textwidth}{!}{
\begin{tabular}{lcccccc}
\hline
 & DESIBAO2 & CMB+DESIBAO2 & PPS & CMB+DESIBAO2+PPS & CC+GRB+Pan & CMB+DESIBAO2+CC+GRB+Pan \\
\hline

$H_0$
& $68.86^{+0.53}_{-0.73}$
& $68.72^{+0.36}_{-0.45}$
& $73.5\pm 1.0$
& $69.31\pm 0.43$
& $67.5\pm 1.7$
& $68.52^{+0.32}_{-0.41}$ \\

$\Omega_{\rm m0}$
& $0.3005^{+0.0087}_{-0.011}$
& $0.3001\pm 0.0040$
& $0.340^{+0.017}_{-0.025}$
& $0.2961\pm 0.0038$
& $0.336^{+0.017}_{-0.021}$
& $0.3023\pm 0.0038$ \\

$\Omega_{\rm b0}$
& $0.0468^{+0.0010}_{-0.00075}$
& $0.04755^{+0.00070}_{-0.00049}$
& $0.0411\pm 0.0015$
& $0.04673\pm 0.00065$
& $0.0488^{+0.0025}_{-0.0029}$
& $0.04782^{+0.00063}_{-0.00041}$ \\

$\Omega_\mathrm{m0} h^2$
& $0.1426^{+0.0045}_{-0.0071}$
& $0.14171^{+0.00087}_{-0.0011}$
& $0.1835^{+0.0095}_{-0.014}$
& $0.1423\pm 0.0011$
& $0.1529^{+0.0081}_{-0.010}$
& $0.14188^{+0.00080}_{-0.0011}$\\

$\Omega_\mathrm{b0} h^2$
& $0.02220\pm 0.00050$
& $0.02242\pm0.00014$
& $0.02220\pm0.00050$
& $0.02245\pm0.00014$
& $0.02222\pm0.00050$
& $0.02245\pm0.00014$ \\

$\log_{10}\xi$
& ---
& ---
& ---
& ---
& ---
& $<0.929$ \\

$\log_{10}(a_i^3 \sqrt{\xi}\kappa\chi_i)$
& $-1.9^{+1.2}$
& $-1.54^{+0.87}$
& $<-0.4$
& $-1.09^{+0.37}_{-0.17}$
& $<-1.31$
& $-1.67^{+0.88}_{-0.29}$ \\

$\mathrm{sign}(\Pi_i)\mathrm{KE}_i/\rho_{\rm DE,0}$
& ---
& $-0.04^{+0.51}_{-0.71}$
& ---
& $-0.04^{+0.42}_{-0.52}$
& ---
& $0.00\pm 0.53$ \\

$(\mathrm{KE}_i + V_*)/\rho_{\rm DE,0}$
& $0.9998^{+0.0033}_{-0.0029}$
& $1.0025^{+0.0021}_{-0.0058}$
& $1.00127^{+0.0047}_{-0.0078}$
& $1.007990^{+0.000072}_{-0.013}$
& $1.0013^{+0.0017}_{-0.0037}$
& $1.0006^{+0.0019}_{-0.0026}$ \\

\hline
$\mathrm{AIC_C}$
  & $34.34\pm0.19$
  & $31.94\pm0.45$
  & $1463.614\pm0.016$
  & $1501.49\pm0.24$
  & $1463.822\pm0.047$
  & $1478.616\pm0.044$ \\
BIC
  & $23.73\pm0.19$
  & $27.24\pm0.45$
  & $1496.199\pm0.016$
  & $1534.13\pm0.24$
  & $1496.792\pm0.047$
  & $1511.639\pm0.044$ \\
DIC
  & $16.00\pm0.15$
  & $22.22\pm0.72$
  & $1456.569\pm0.053$
  & $1500.608\pm0.027$
  & $1458.191\pm0.066$
  & $1477.94\pm0.35$ \\
WAIC
  & $14.310\pm0.099$
  & $20.84\pm0.41$
  & $1456.18\pm0.11$
  & $1500.97\pm0.38$
  & $1457.884\pm0.090$
  & $1476.68\pm0.31$ \\
$-\ln{\rm B}$
  & $16.31\pm0.67$
  & $21.53\pm0.13$
  & $1457.17\pm0.62$
  & $1501.505\pm0.058$
  & $1457.97\pm0.12$
  & $1476.94\pm0.40$ \\

\hline
$\Delta\mathrm{AIC_C}$
  & $15.39\pm0.19$
  & $8.67\pm0.45$
  & $5.583\pm0.016$
  & $-2.68\pm0.24$
  & $5.860\pm0.047$
  & $1.068\pm0.044$ \\
$\Delta\mathrm{BIC}$
  & $5.75\pm0.19$
  & $3.66\pm0.45$
  & $21.865\pm0.016$
  & $13.63\pm0.24$
  & $22.335\pm0.047$
  & $17.570\pm0.044$ \\
$\Delta\mathrm{DIC}$
  & $1.79\pm0.32$
  & $1.42\pm0.74$
  & $0.49\pm0.13$
  & $-3.148\pm0.080$
  & $0.39\pm0.46$
  & $0.42\pm0.50$ \\
$\Delta\mathrm{WAIC}$
  & $0.19\pm0.31$
  & $0.20\pm0.52$
  & $0.15\pm0.17$
  & $-2.55\pm0.42$
  & $-0.05\pm0.99$
  & $-1.0\pm1.0$ \\
$\Delta(-\ln{\rm B})$
  & $2.00\pm0.88$
  & $-0.6\pm2.4$
  & $1.21\pm0.63$
  & $-1.77\pm0.52$
  & $-1.3\pm1.7$
  & $-2.2\pm3.5$ \\
\hline
\end{tabular}
}
\caption{\justifying{Same as Table~\ref{tab:LCDM_params} for the three-form model, including the $\Delta$IC values when compared with $\Lambda$CDM. Entries are reported with a central value and only an upper bound when the $95\%$ interval only provides an upper bound. Entries marked ``---'' indicate that the combination provides no constraint beyond the prior.}}\label{tab:3form_params}
\end{table*}

\subsection{Model comparison and dataset consistency}\label{sec:ic_tension}

\subsubsection{Information criteria}\label{sec:IC}

The differences in information criteria $\Delta\mathrm{IC}\equiv\mathrm{IC}_\mathrm{3\text{-}form}-\mathrm{IC}_{\Lambda\mathrm{CDM}}$ are presented in table~\ref{tab:3form_params}.

For all dataset combinations that are only early (DESIBAO2 and CMB+DESIBAO2) or late (PPS and CC+GRB+Pan) data, every criterion favours $\Lambda$CDM or is inconclusive within uncertainties. This is coherent with the previous observations; when fitted to these datasets in isolation, many of the three-form parameters are prior-dominated and contribute model complexity without a compensating improvement in fit quality.

For the combination CMB+DESIBAO2+CC+GRB +Pan, $\mathrm{AIC_C}$ and the Bayesian approach criteria are still inconclusive, but we can observe that in general these criteria now yield values considerably lower compared to the isolated dataset combinations, suggesting that the better fit of the three-form model to these data can start compensating its extra parameters compared to  $\Lambda$CDM.

Furthermore, for the combined dataset CMB+\allowbreak{}DESIBAO2+\allowbreak{}PPS, $\mathrm{AIC_C}$ and the Bayesian criteria  all change sign. In this case, the detection for preference for the three-form model is moderate and well-established within uncertainties.

Note that the Bayesian evidence criterion values tend to have large uncertainties that come from its definition being dominated by outlier terms in the average.

The BIC remains positive for all combined fits, favouring $\Lambda$CDM. This is mostly expected, as for the small CMB and BAO datasets the three-form model does not provide much fit improvement, while for large-$n$ datasets (SNe), the $k\ln n$ penalty term is large. 

\subsubsection{Dataset consistency and Hubble tension}\label{sec:tension}

The goodness of fit (GoF) and suspiciousness (S) tension probes defined in subsection~\ref{sec:model-comparison} are evaluated for two dataset pairs and reported in table~\ref{tab:tension}: early-time CMB+DESIBAO2 against both late-time combinations, PPS and CC+GRB+Pan.

For pair 1, both models show a strong tension, driven by the Hubble tension, the $\sim5$\,km\,s$^{-1}$\,Mpc$^{-1}$ separation of the CMB+DESIBAO2 and PPS $H_0$ posteriors. The three-form model cannot reduce this tension for either probe. While it allows higher values of $H_0$ in dataset combinations that include CMB (see figure~\ref{fig:contour_pps}), and can achieve lower GoF and S values than $\Lambda$CDM, the $n-\sigma$ conversion also depends on the Bayesian model dimension ${\rm BMD}_{\rm eff}$. This ${\rm BMD}_{\rm eff}$ is smaller for the three-form model in this dataset pair, which indicates that the parameter directions that are unconstrained in the individual fits become constrained in the combined fit to accommodate the tension, raising ${\rm BMD} (D_1D_2|M)$ above both individual values ${\rm BMD} (D_1|M)$ and ${\rm BMD} (D_2|M)$. The resulting steeper mapping to $\sigma$ values compensates the reduction in GoF and S values.

For pair 2, both models find CMB+DESIBAO2 and CC+GRB+Pan to be consistent with a maximum separation of $\approx 1.6\sigma$ ($\mathrm{\Lambda}$CDM GoF). In this case the three-form model does show a weak but robust tension  reduction in S, while the GoF reduction is marginal.

\begin{table*}
\centering
\begin{tabular}{llcc}
\hline
Probe &  & CMB+DESIBAO2 $\sim$ PPS & CMB+DESIBAO2 $\sim$ CC+GRB+Pan \\
\hline
\multirow{3}{*}{$n_\sigma(\mathrm{GoF})$}
  & $\Lambda$CDM & $5.25532\pm0.00087$ & $1.6225\pm0.0021$ \\
  & three-form   & $5.351\pm0.053$     & $1.516\pm0.099$ \\
  & $\Delta$     & $+0.096\pm0.053$    & $-0.107\pm0.099$ \\
\hline
\multirow{3}{*}{$n_\sigma(S)$}
  & $\Lambda$CDM & $5.068\pm0.011$  & $0.901\pm0.097$ \\
  & three-form   & $5.085\pm0.025$  & $0.597\pm0.074$ \\
  & $\Delta$     & $+0.017\pm0.027$ & $-0.30\pm0.12$ \\
\hline
\end{tabular}
\caption{\justifying{Dataset consistency tension probes, expressed in $n_\sigma$ units for both models (see subsection ~\ref{sec:model-comparison}). Positive $n_\sigma$: tension; negative $n_\sigma(S)$: datasets are more complementary than expected. Negative $\Delta$ values indicate the three-form model reduces tension relative to $\Lambda$CDM.}}
\label{tab:tension}
\end{table*}

Taken together, the IC and tension results point towards the three-form's extra degrees of freedom being potentially useful in the fits that combine early and late time data. However, after applying the appropriate statistical penalties to these  degrees of freedom, the result is in most cases a fit of comparable quality to $\Lambda$CDM. This  makes it difficult to justify the additional complexity in this specific three-form model with the currently available data. The only exception being the clear if mild preference for the three-form over $\Lambda$CDM expressed by the $\mathrm{AIC_C}$, DIC, WAIC, and $-\ln\mathrm{B}$ in the CMB+DESIBAO2+PPS case.

Comparing with the previous results of \cite{Bouhmadi-Lopez:2025lzm} in the dataset combinations that are comparable, such as DESIBAO and PPS (comparable to  SNe+low-z in that paper), we can see that the values obtained for the physical quantities such as $H_0$, $\Omega_{m0} h^2$ and $\Omega_{b0}h^2$ are consistent. As for the statistical quantities, they are also consistent up to an extra factor of 2 in their definitions (see footnote~\ref{foot:2factor}).

\subsection{Dark energy evolution}\label{sec:evolution}

\begin{figure*}
\centering
\begin{subfigure}[t]{0.72\textwidth}
    \centering
    \includegraphics[width=\linewidth]{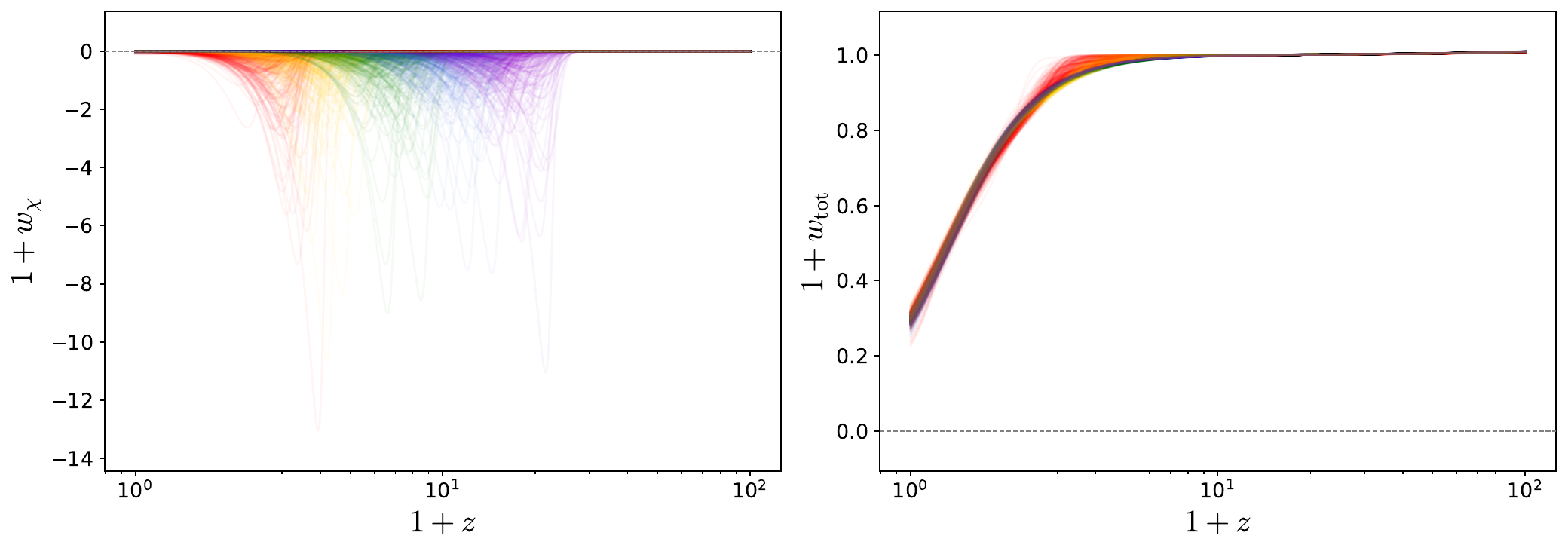}
    \caption{DESIBAO2}
    \label{fig:evol_bao_only}
\end{subfigure}
\\[0pt]
\begin{subfigure}[t]{0.72\textwidth}
    \centering
    \includegraphics[width=\linewidth]{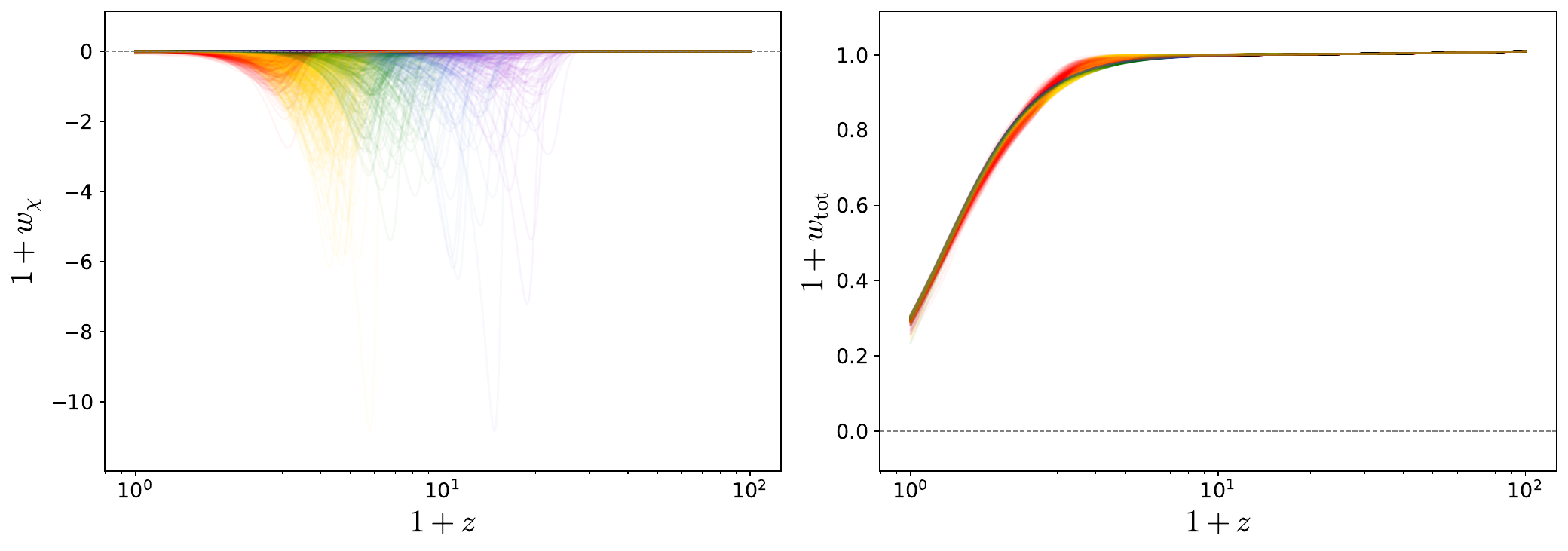}
    \caption{CMB+DESIBAO2}
    \label{fig:evol_bao}
\end{subfigure}
\\[0pt]
\begin{subfigure}[t]{0.72\textwidth}
    \centering
    \includegraphics[width=\linewidth]{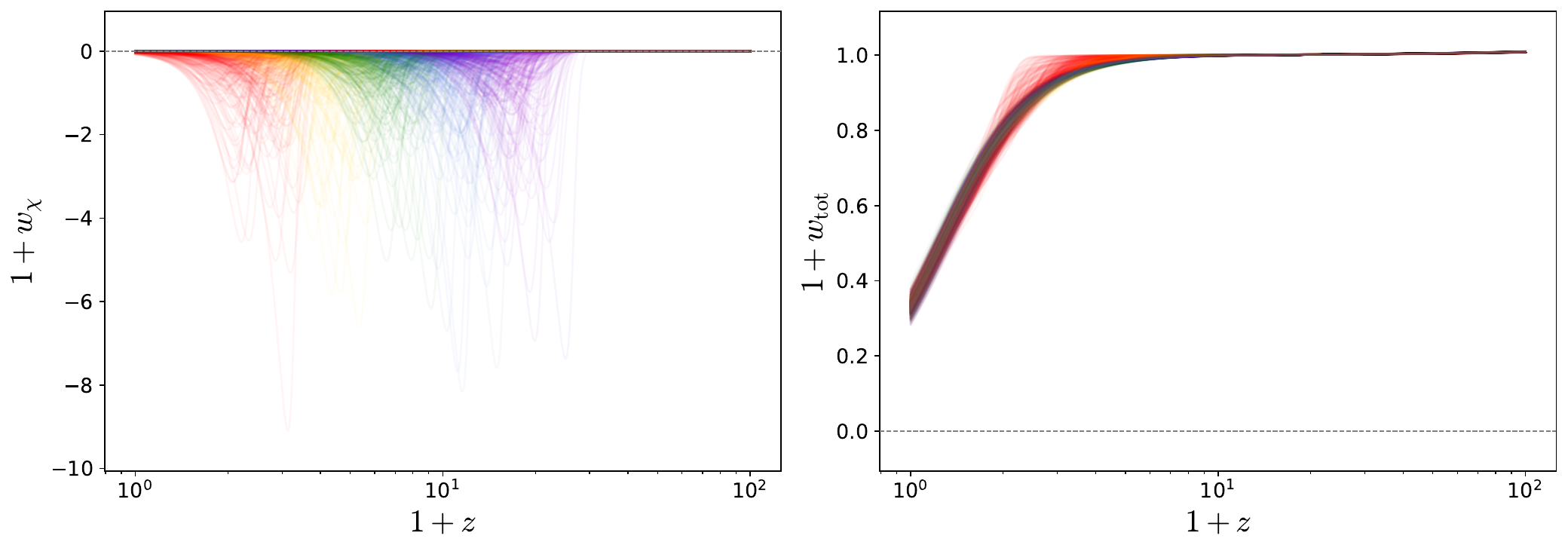}
    \caption{CC+GRB+Pan}
    \label{fig:evol_cc}
\end{subfigure}
\\[0pt]
\begin{subfigure}[t]{0.72\textwidth}
    \centering
    \includegraphics[width=\linewidth]{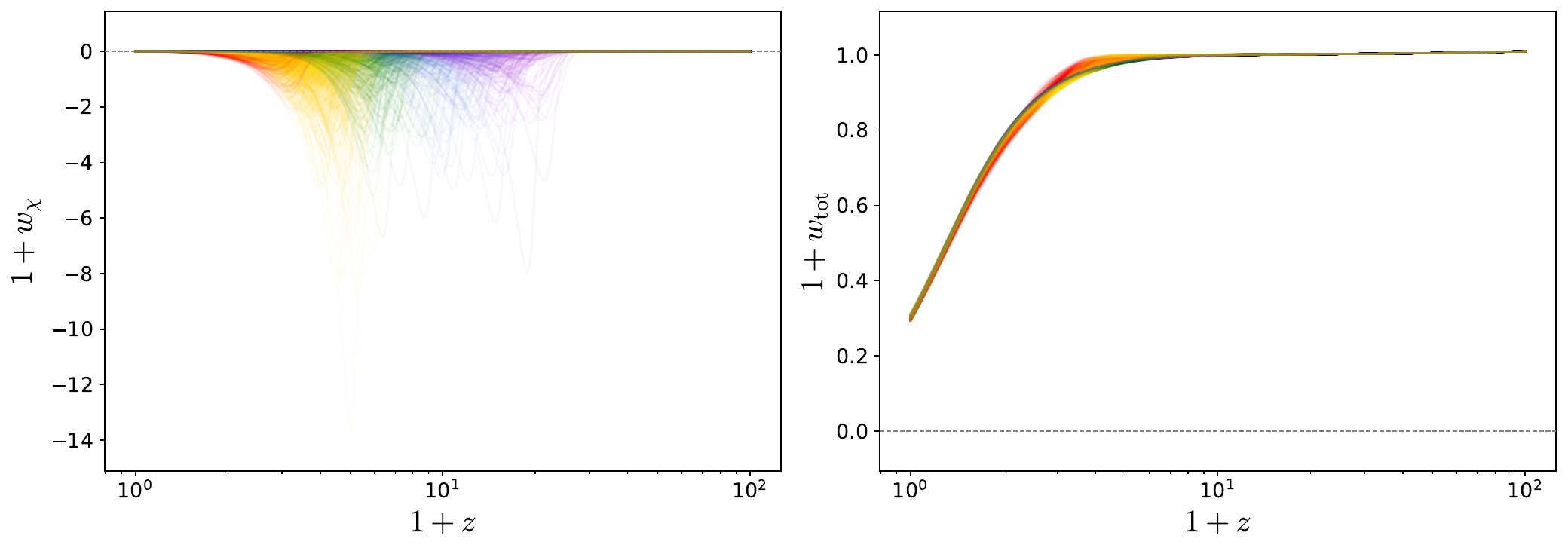}
    \caption{CMB+DESIBAO2+CC+GRB+Pan}
    \label{fig:evol_ccc}
\end{subfigure}
\caption{\justifying{Three-form dark energy equation-of-state $1+w_\chi$ (\textit{left}) and total equation-of-state $1+w_\mathrm{tot}$ (\textit{right}) as functions of redshift for 400 posterior samples. Colour indicates $\log_{10}(a_i^3\sqrt{\xi}\kappa\chi_i)$: red $>-1.0$, yellow $-1.625$ to $-1.0$, green $-2.25$ to $-1.625$, blue $-2.875$ to $-2.25$, violet $<-2.875$}}
\label{fig:w_evolution}
\end{figure*}

Taking random samples from the MCMC chains, we can obtain a representative evolution of parameters for the three-form model. In particular, figure~\ref{fig:w_evolution} shows the evolution of the three-form dark energy and total equation of state (EOS) $w_\mathrm{\chi}(z)$ and $w_{tot}$ for the two early-time dataset fits: DESIBAO2 and CMB+DESIBAO2, as well as the late-time representative CC+GRB+Pan and the full combination of these, CMB+DESIBAO2+CC+GRB+Pan, colour-coded by the initial field amplitude $\log_{10}(a_i^3\sqrt{\xi}\kappa\chi_i)$.

In all cases it can be observed that, at first, while the three-form rolls up the potential far from the top, its EOS $w_{\chi}$ remains very close to $-1$. Then, there is a time where it has a stronger phantom behaviour, dipping considerably below $-1$. Finally, the EOS returns towards $-1$, and it will keep approaching that value asymptotically as it moves towards the LSBR attractor. All in all, the three-form behaves close to a cosmological constant in the asymptotic past and future, but has phantom behaviour at intermediate redshifts.

From the colours we can also deduce that a larger initial field makes the dip happen at lower redshift, so the fact that the early-time datasets CMB and DESIBAO2 impose a tighter upper limit on this parameter (as seen in figure~\ref{fig:scatter_bao}) also restricts the latest time at which the dip can happen. This is also the case for the combinations that include these datasets such as CMB+DESIBAO2+CC+GRB+Pan. Meanwhile, late-time only datasets alone do not constrain this parameter nor the time of this phantom dip very strongly.

\section{Conclusions}\label{conclusions}

Three-form fields constitute an appealing framework for describing dark energy and the late-time accelerated expansion of the Universe. Their theoretical motivation stems from higher-dimensional constructions, where antisymmetric tensor fields naturally emerge, while their cosmological dynamics allow for a wide variety of behaviours depending on the underlying potential and interactions. In particular, three-form cosmologies can reproduce effective cosmological constant scenarios, quintessence-like evolution, or phantom dynamics, thus providing a flexible setting for modelling the current accelerated era. Beyond dark energy, these fields have also found applications in several areas of gravitational and cosmological physics, including inflationary dynamics, compact objects, wormhole geometries, modified gravity scenarios, anisotropic cosmologies, and quantum cosmology. Such versatility reinforces the relevance of three-form theories as promising candidates for extending the standard cosmological framework.

Within this setup, we explore a DE scenario driven by a minimally coupled three-form field characterised by a Gaussian potential. The dynamics of the field give rise to accelerated cosmic expansion and admit a broad range of late-time behaviours, spanning from effective cosmological-constant solutions to phantom-like regimes. By performing a background-level comparison with current observational data, we constrain the parameter space of the model and evaluate its consistency as a viable alternative to the standard description of late-time acceleration. In particular, the data we have used  correspond to the CMB shift parameters, the DESI DR2 baryon acoustic oscillation measurements, the Pantheon+ Type Ia supernova catalogue both with and without the SH0ES distance-ladder calibration, cosmic chronometers, and gamma-ray bursts, combined into six dataset combinations spanning the early-time, distance-ladder high-$H_0$, and low-$H_0$ groups. The observational fits have been carried out within a Bayesian Markov-chain Monte Carlo framework, implemented through the \texttt{Cobaya} package. We have also compared the fits among different dataset combinations through several information criteria: the corrected Akaike ($\mathrm{AIC_C}$), Bayesian ($\mathrm{BIC}$), deviance ($\mathrm{DIC}$), and widely-applicable ($\mathrm{WAIC}$) information criteria, together with the Bayes evidence ($\ln \mathrm{B}$). We have also quantified the tension between dataset pairs using the goodness-of-fit ($\mathrm{GoF}$) and suspiciousness ($\mathrm{S}$) statistics for both the three-form model and the standard $\mathrm{\Lambda}$CDM model.

Our analysis implies that, while the three-form dark energy model, with its phantom dip and asymptotic cosmological constant-like behaviour, is a viable description of the background history of the Universe, the information criteria favour $\Lambda$CDM for all dataset combinations restricted to either late or early time data, where the extra parameters of the three-form are not strongly constrained and contribute model complexity without a compensating improvement in fit.
This trend reverses progressively as more datasets are combined together. Aside from BIC, for the combination CMB+DESIBAO2+CC+GRB+Pan the rest of the criteria become neutral, and turn slightly in favour of the three-form for the combined CMB+DESIBAO2+PPS dataset, indicating a moderate preference for the phantom-like three-form model. This shows that the extra degrees of freedom only become useful when fitting data of different origins. While the statistical evidence is not conclusive, it points to a possible improvement in describing a dataset combination that is in tension under $\Lambda$CDM.

The tension analysis reveals that the CMB+DESIBAO2 versus PPS discrepancy is high, above $5\sigma$ for both models and tension probes; the three-form model does not alleviate it even with its broader $H_0$ posterior. Even with a reduction in the value of the raw S and GoF in the three-form case, this gain is offset by the lower effective Bayesian model dimension $\mathrm{BMD}_{\rm eff}$: parameter directions left unconstrained in the individual fits become constrained once the joint fit must accommodate the tension, which steepens the mapping to $n_\sigma$ and cancels out the improvement. 
This inability of the three-form model to reduce the tension even with its higher $H_0$ values highlights the nature of the Hubble tension, i.e. the direct observational constraint of $\Omega_{\rm m0} h^2$ by local distance measurements in conflict with the observed acoustic peak of CMB and matter power spectrum. For datasets of similar origins, the CMB+DESIBAO2 and CC+GRB+Pan datasets are found to be marginally more consistent in the three-form case than under $\Lambda$CDM, with a robust reduction in suspiciousness and a marginal reduction in the goodness of fit.

The reconstructed evolution of the three-form equation of state reveals a common qualitative pattern across posterior samples: $w_{\rm DE}$ remains close to
$-1$ in the asymptotic past and future, approaching the
LSBR attractor at late times, while exhibiting a phantom dip at intermediate redshifts that separates the cosmological dynamics from those of $\rm\Lambda$CDM, and whose timing is set by the initial field amplitude $\log_{10}(a_i^3\sqrt{\xi}\kappa\chi_i)$. Early-time data (CMB, DESIBAO2) bound this amplitude most strongly from above and thus restrict how late the dip can occur.


In conclusion, three-form dark energy remains a theoretically rich and observationally viable description of late-time cosmic acceleration. For the specific Gaussian potential handled in this work, there is a small statistical preference over $\rm \Lambda$CDM when early-time and late-time high-$H_0$ data are combined, though it does not by itself alleviate the Hubble tension. The model therefore stands as a well-motivated, if not yet decisively preferred, alternative to $\Lambda$CDM. Future data  could help discriminate the viability of the specific dynamical features of this Gaussian three-form model, most notably the phantom dip.
\section*{Acknowledgements}

 M. B.-L. is supported by the Basque Foundation for Science Ikerbasque.  H.W.C is supported by the NSFC Grant No.~12250410250, No.~12505074 and No.~12347133.  C. G. B. acknowledges financial support from the FPI fellowship PRE2021-100340 of the Spanish Ministry of Science, Innovation and Universities. J.O.d.R.  is supported by the FPI fellowship PREP2023-002243 of the Spanish Ministry of Science, Innovation and Universities.
 T. J. B.  is supported by the Basque Foundation for Science Ikerbasque.
 Our work is supported by the Spanish Grant PID2023-149016NB-I00 (MINECO/AEI/FEDER, UE). This work is also supported by the Basque Government grant No. IT1977-26 (Spain). The authors acknowledge the contribution of the COST Action CA21136 “Addressing observational tensions in cosmology with systematics and fundamental physics (CosmoVerse)”.
\bibliographystyle{elsarticle-num}
\bibliography{merged_bibliography}

@article{Ibarra-Uriondo:2026zbp,
    author = "Ibarra-Uriondo, Be{\~n}at and Bouhmadi-L{\'o}pez, Mariam",
    title = "{Sign-switching dark energy: Smooth transitions with recent DESI DR2 observations}",
    eprint = "2602.12347",
    archivePrefix = "arXiv",
    primaryClass = "astro-ph.CO",
    doi = "10.1016/j.dark.2026.102351",
    journal = "Phys. Dark Univ.",
    volume = "52",
    pages = "102351",
    year = "2026"
}

@article{Bento:2002ps,
    author = "Bento, M. C. and Bertolami, O. and Sen, A. A.",
    title = "{Generalized Chaplygin gas, accelerated expansion and dark energy matter unification}",
    eprint = "gr-qc/0202064",
    archivePrefix = "arXiv",
    doi = "10.1103/PhysRevD.66.043507",
    journal = "Phys. Rev. D",
    volume = "66",
    pages = "043507",
    year = "2002"
}

@article{DiValentino:2026uua,
    author = "Di Valentino, Eleonora",
    title = "{Cracks in the Standard Cosmological Model: Anomalies, Tensions, and Hints of New Physics}",
    eprint = "2601.01525",
    archivePrefix = "arXiv",
    primaryClass = "astro-ph.CO",
    doi = "10.22323/1.507.0012",
    journal = "PoS",
    volume = "COSMICWISPers2025",
    pages = "012",
    year = "2026"
}

@article{Bouhmadi-Lopez:2006fwq,
    author = "Bouhmadi-L\'opez, Mariam and Gonz\'{a}lez-D\'{i}az, Pedro F. and Mart\'in-Moruno, Prado",
    title = "{Worse than a big rip?}",
    eprint = "gr-qc/0612135",
    archivePrefix = "arXiv",
    doi = "10.1016/j.physletb.2007.10.079",
    journal = "Phys. Lett. B",
    volume = "659",
    pages = "1--5",
    year = "2008"
}

@article{Linder:2002et,
    author = "Linder, Eric V.",
    title = "{Exploring the expansion history of the universe}",
    eprint = "astro-ph/0208512",
    archivePrefix = "arXiv",
    doi = "10.1103/PhysRevLett.90.091301",
    journal = "Phys. Rev. Lett.",
    volume = "90",
    pages = "091301",
    year = "2003"
}

@article{Chevallier:2000qy,
    author = "Chevallier, Michel and Polarski, David",
    title = "{Accelerating universes with scaling dark matter}",
    eprint = "gr-qc/0009008",
    archivePrefix = "arXiv",
    doi = "10.1142/S0218271801000822",
    journal = "Int. J. Mod. Phys. D",
    volume = "10",
    pages = "213--224",
    year = "2001"
}

@article{Copeland:2006wr,
    author = "Copeland, Edmund J. and Sami, M. and Tsujikawa, Shinji",
    title = "{Dynamics of dark energy}",
    eprint = "hep-th/0603057",
    archivePrefix = "arXiv",
    doi = "10.1142/S021827180600942X",
    journal = "Int. J. Mod. Phys. D",
    volume = "15",
    pages = "1753--1936",
    year = "2006"
}

@article{Verde:2019ivm,
    author = "Verde, L. and Treu, T. and Riess, A. G.",
    title = "{Tensions between the Early and the Late Universe}",
    eprint = "1907.10625",
    archivePrefix = "arXiv",
    primaryClass = "astro-ph.CO",
    doi = "10.1038/s41550-019-0902-0",
    journal = "Nature Astron.",
    volume = "3",
    pages = "891",
    year = "2019"
}

@article{Riess:2019cxk,
    author = "Riess, Adam G. and Casertano, Stefano and Yuan, Wenlong and Macri, Lucas M. and Scolnic, Dan",
    title = "{Large Magellanic Cloud Cepheid Standards Provide a 1\% Foundation for the Determination of the Hubble Constant and Stronger Evidence for Physics beyond {$\Lambda$CDM}}",
    eprint = "1903.07603",
    archivePrefix = "arXiv",
    primaryClass = "astro-ph.CO",
    doi = "10.3847/1538-4357/ab1422",
    journal = "Astrophys. J.",
    volume = "876",
    number = "1",
    pages = "85",
    year = "2019"
}

@article{Koivisto:2009fb,
    author = "Koivisto, Tomi S. and Nunes, Nelson J.",
    title = "{Inflation and dark energy from three-forms}",
    eprint = "0908.0920",
    archivePrefix = "arXiv",
    primaryClass = "astro-ph.CO",
    doi = "10.1103/PhysRevD.80.103509",
    journal = "Phys. Rev. D",
    volume = "80",
    pages = "103509",
    year = "2009"
}

@article{Bouhmadi-Lopez:2025lzm,
    author = "Bouhmadi-L{\'o}pez, Mariam and Chiang, Hsu-Wen and Boiza, Carlos G. and Chen, Pisin",
    title = "{Observational constraints on 3-forms dark energy}",
    eprint = "2512.09991",
    archivePrefix = "arXiv",
    primaryClass = "astro-ph.CO",
    doi = "10.1088/1475-7516/2026/06/084",
    journal = "JCAP",
    volume = "06",
    pages = "084",
    year = "2026"
}

@article{Morais:2016bev,
    author = "Morais, Jo{\~a}o and Bouhmadi-L{\'o}pez, Mariam and Sravan Kumar, K. and Marto, Jo{\~a}o and Tavakoli, Yaser",
    title = "{Interacting 3-form dark energy models: distinguishing interactions and avoiding the Little Sibling of the Big Rip}",
    eprint = "1608.01679",
    archivePrefix = "arXiv",
    primaryClass = "gr-qc",
    doi = "10.1016/j.dark.2016.11.002",
    journal = "Phys. Dark Univ.",
    volume = "15",
    pages = "7--30",
    year = "2017"
}

@article{Bouhmadi-Lopez:2016dzw,
    author = "Bouhmadi-L{\'o}pez, Mariam and Marto, Jo{\~a}o and Morais, Jo{\~a}o and Silva, C{\'e}sar M.",
    title = "{Cosmic infinity: A dynamical system approach}",
    eprint = "1611.03100",
    archivePrefix = "arXiv",
    primaryClass = "gr-qc",
    doi = "10.1088/1475-7516/2017/03/042",
    journal = "JCAP",
    volume = "03",
    pages = "042",
    year = "2017"
}

@article{Akarsu:2025nns,
    author = {Akarsu, {\"O}zg{\"u}r and Di Valentino, Eleonora and Vysko{\v{c}}il, Ji{\v{r}}{\'\i} and Y{\i}lmaz, Ezgi and Y{\"u}kselci, A. Emrah and Zhuk, Alexander},
    title = "{Nonlinear matter power spectrum from relativistic N-body simulations: {\ensuremath{\Lambda}}sCDM versus {\ensuremath{\Lambda}}CDM}",
    eprint = "2510.18741",
    archivePrefix = "arXiv",
    primaryClass = "astro-ph.CO",
    doi = "10.1103/lmt4-hshz",
    journal = "Phys. Rev. D",
    volume = "113",
    number = "8",
    pages = "083508",
    year = "2026"
}

@article{CosmoVerseNetwork:2025alb,
    author = "Di Valentino, Eleonora and others",
    collaboration = "CosmoVerse Network",
    title = "{The CosmoVerse White Paper: Addressing observational tensions in cosmology with systematics and fundamental physics}",
    eprint = "2504.01669",
    archivePrefix = "arXiv",
    primaryClass = "astro-ph.CO",
    doi = "10.1016/j.dark.2025.101965",
    journal = "Phys. Dark Univ.",
    volume = "49",
    pages = "101965",
    year = "2025"
}

@article{Ratra:1987rm,
    author = "Ratra, Bharat and Peebles, P. J. E.",
    title = "{Cosmological Consequences of a Rolling Homogeneous Scalar Field}",
    reportNumber = "PUPT-1072",
    doi = "10.1103/PhysRevD.37.3406",
    journal = "Phys. Rev. D",
    volume = "37",
    pages = "3406",
    year = "1988"
}

@article{Caldwell1998,
    author = "Caldwell, Robert R. and Dave, Rahul and Steinhardt, Paul J.",
    title = "{Cosmological imprint of an energy component with general equation of state}",
    journal = "Phys. Rev. Lett.",
    volume = "80",
    pages = "1582--1585",
    year = "1998",
    doi = "10.1103/PhysRevLett.80.1582"
}

@article{Planck2018,
    author = "{Planck Collaboration}",
    title = "{Planck 2018 results. VI. Cosmological parameters}",
    journal = "Astron. Astrophys.",
    volume = "641",
    pages = "A6",
    year = "2020",
    doi = "10.1051/0004-6361/201833910"
}

@article{Chiang:2025qxg,
    author = "Chiang, Hsu-Wen and Boiza, Carlos G. and Bouhmadi-L{\'o}pez, Mariam",
    title = "{Observational constraints on generalised axion-like potentials for the late Universe}",
    eprint = "2503.04898",
    archivePrefix = "arXiv",
    primaryClass = "astro-ph.CO",
    doi = "10.1088/1475-7516/2025/08/064",
    journal = "JCAP",
    volume = "08",
    pages = "064",
    year = "2025"
}

@article{SupernovaSearchTeam:1998fmf,
    author = "Riess, Adam G. and others",
    collaboration = "Supernova Search Team",
    title = "{Observational evidence from supernovae for an accelerating universe and a cosmological constant}",
    eprint = "astro-ph/9805201",
    archivePrefix = "arXiv",
    doi = "10.1086/300499",
    journal = "Astron. J.",
    volume = "116",
    pages = "1009--1038",
    year = "1998"
}

@article{SupernovaCosmologyProject:1998vns,
    author = "Perlmutter, S. and others",
    collaboration = "Supernova Cosmology Project",
    title = "{Measurements of {$\Omega$} and {$\Lambda$} from 42 High Redshift Supernovae}",
    eprint = "astro-ph/9812133",
    archivePrefix = "arXiv",
    reportNumber = "LBNL-41801, LBL-41801",
    doi = "10.1086/307221",
    journal = "Astrophys. J.",
    volume = "517",
    pages = "565--586",
    year = "1999"
}

@article{Weinberg:1988cp,
    author = "Weinberg, Steven",
    editor = "Hsu, Jong-Ping and Fine, D.",
    title = "{The Cosmological Constant Problem}",
    reportNumber = "UTTG-12-88",
    doi = "10.1103/RevModPhys.61.1",
    journal = "Rev. Mod. Phys.",
    volume = "61",
    pages = "1--23",
    year = "1989"
}

@article{Padmanabhan:2002ji,
    author = "Padmanabhan, T.",
    title = "{Cosmological constant: The Weight of the vacuum}",
    eprint = "hep-th/0212290",
    archivePrefix = "arXiv",
    doi = "10.1016/S0370-1573(03)00120-0",
    journal = "Phys. Rept.",
    volume = "380",
    pages = "235--320",
    year = "2003"
}

@article{Kamenshchik:2001cp,
    author = "Kamenshchik, Alexander Yu. and Moschella, Ugo and Pasquier, Vincent",
    title = "{An Alternative to quintessence}",
    eprint = "gr-qc/0103004",
    archivePrefix = "arXiv",
    doi = "10.1016/S0370-2693(01)00571-8",
    journal = "Phys. Lett. B",
    volume = "511",
    pages = "265--268",
    year = "2001"
}

@book{CANTATA:2021asi,
    author = "Akrami, Yashar and others",
    editor = "Saridakis, Emmanuel N. and Lazkoz, Ruth and Salzano, Vincenzo and Vargas Moniz, Paulo and Capozziello, Salvatore and Beltr\'an Jim\'enez, Jose and De Laurentis, Mariafelicia and Olmo, Gonzalo J.",
    collaboration = "CANTATA",
    title = "{Modified Gravity and Cosmology. An Update by the CANTATA Network}",
    eprint = "2105.12582",
    archivePrefix = "arXiv",
    primaryClass = "gr-qc",
    doi = "10.1007/978-3-030-83715-0",
    isbn = "978-3-030-83714-3, 978-3-030-83717-4, 978-3-030-83715-0",
    publisher = "Springer",
    year = "2021"
}

@article{Bouhmadi-Lopez:2014cca,
    author = "Bouhmadi-L\'opez, Mariam and Errahmani, Ahmed and Mart\'in-Moruno, Prado and Ouali, Taoufik and Tavakoli, Yaser",
    title = "{The little sibling of the big rip singularity}",
    eprint = "1407.2446",
    archivePrefix = "arXiv",
    primaryClass = "gr-qc",
    doi = "10.1142/S0218271815500789",
    journal = "Int. J. Mod. Phys. D",
    volume = "24",
    number = "10",
    pages = "1550078",
    year = "2015"
}

@article{Morais:2015ooa,
    author = "Morais, Jo{\~a}o and Bouhmadi-L{\'o}pez, Mariam and Capozziello, Salvatore",
    title = "{Can {$f(R)$} gravity contribute to (dark) radiation?}",
    eprint = "1507.02623",
    archivePrefix = "arXiv",
    primaryClass = "gr-qc",
    doi = "10.1088/1475-7516/2015/09/041",
    journal = "JCAP",
    volume = "09",
    pages = "041",
    year = "2015"
}

@article{Capozziello:2011et,
    author = "Capozziello, Salvatore and De Laurentis, Mariafelicia",
    title = "{Extended Theories of Gravity}",
    eprint = "1108.6266",
    archivePrefix = "arXiv",
    primaryClass = "gr-qc",
    doi = "10.1016/j.physrep.2011.09.003",
    journal = "Phys. Rept.",
    volume = "509",
    pages = "167--321",
    year = "2011"
}

@article{Nojiri:2010wj,
    author = "Nojiri, Shin'ichi and Odintsov, Sergei D.",
    title = "{Unified cosmic history in modified gravity: from F(R) theory to Lorentz non-invariant models}",
    eprint = "1011.0544",
    archivePrefix = "arXiv",
    primaryClass = "gr-qc",
    doi = "10.1016/j.physrep.2011.04.001",
    journal = "Phys. Rept.",
    volume = "505",
    pages = "59--144",
    year = "2011"
}

@article{Nojiri:2017ncd,
    author = "Nojiri, S. and Odintsov, S. D. and Oikonomou, V. K.",
    title = "{Modified Gravity Theories on a Nutshell: Inflation, Bounce and Late-time Evolution}",
    eprint = "1705.11098",
    archivePrefix = "arXiv",
    primaryClass = "gr-qc",
    reportNumber = "PHYS.REPT.-692-(2017)-1-104, Phys.Rept. 692 (2017) 1-104",
    doi = "10.1016/j.physrep.2017.06.001",
    journal = "Phys. Rept.",
    volume = "692",
    pages = "1--104",
    year = "2017"
}

@article{Bengochea:2008gz,
    author = "Bengochea, Gabriel R. and Ferraro, Rafael",
    title = "{Dark torsion as the cosmic speed-up}",
    eprint = "0812.1205",
    archivePrefix = "arXiv",
    primaryClass = "astro-ph",
    doi = "10.1103/PhysRevD.79.124019",
    journal = "Phys. Rev. D",
    volume = "79",
    pages = "124019",
    year = "2009"
}

@article{Ferraro:2006jd,
    author = "Ferraro, Rafael and Fiorini, Franco",
    title = "{Modified teleparallel gravity: Inflation without inflaton}",
    eprint = "gr-qc/0610067",
    archivePrefix = "arXiv",
    doi = "10.1103/PhysRevD.75.084031",
    journal = "Phys. Rev. D",
    volume = "75",
    pages = "084031",
    year = "2007"
}

@article{Cai:2015emx,
    author = "Cai, Yi-Fu and Capozziello, Salvatore and De Laurentis, Mariafelicia and Saridakis, Emmanuel N.",
    title = "{f(T) teleparallel gravity and cosmology}",
    eprint = "1511.07586",
    archivePrefix = "arXiv",
    primaryClass = "gr-qc",
    doi = "10.1088/0034-4885/79/10/106901",
    journal = "Rept. Prog. Phys.",
    volume = "79",
    number = "10",
    pages = "106901",
    year = "2016"
}

@article{BeltranJimenez:2018vdo,
    author = "Beltr{\'a}n Jim{\'e}nez, Jose and Heisenberg, Lavinia and Koivisto, Tomi S.",
    title = "{Teleparallel Palatini theories}",
    eprint = "1803.10185",
    archivePrefix = "arXiv",
    primaryClass = "gr-qc",
    reportNumber = "NORDITA-2018-023, IFT-UAM/CSIC-18-035, IFT-UAM-CSIC-18-035",
    doi = "10.1088/1475-7516/2018/08/039",
    journal = "JCAP",
    volume = "08",
    pages = "039",
    year = "2018"
}

@article{BeltranJimenez:2019tme,
    author = "Beltr{\'a}n Jim{\'e}nez, Jose and Heisenberg, Lavinia and Koivisto, Tomi Sebastian and Pekar, Simon",
    title = "{Cosmology in $f(Q)$ geometry}",
    eprint = "1906.10027",
    archivePrefix = "arXiv",
    primaryClass = "gr-qc",
    doi = "10.1103/PhysRevD.101.103507",
    journal = "Phys. Rev. D",
    volume = "101",
    number = "10",
    pages = "103507",
    year = "2020"
}

@article{Ayuso:2020dcu,
    author = "Ayuso, Ismael and Lazkoz, Ruth and Salzano, Vincenzo",
    title = "{Observational constraints on cosmological solutions of $f(Q)$ theories}",
    eprint = "2012.00046",
    archivePrefix = "arXiv",
    primaryClass = "astro-ph.CO",
    doi = "10.1103/PhysRevD.103.063505",
    journal = "Phys. Rev. D",
    volume = "103",
    number = "6",
    pages = "063505",
    year = "2021"
}

@article{Boiza:2025xpn,
    author = "Boiza, Carlos G. and Petronikolou, Maria and Bouhmadi-L{\'o}pez, Mariam and Saridakis, Emmanuel N.",
    title = "{Addressing H $_{0}$ and S $_{8}$ tensions within f(Q) cosmology}",
    eprint = "2505.18264",
    archivePrefix = "arXiv",
    primaryClass = "astro-ph.CO",
    doi = "10.1088/1475-7516/2025/12/011",
    journal = "JCAP",
    volume = "12",
    pages = "011",
    year = "2025"
}

@article{Ayuso:2025vkc,
    author = "Ayuso, Ismael and Bouhmadi-L{\'o}pez, Mariam and Chen, Che-Yu and Chew, Xiao Yan and Dialektopoulos, Konstantinos and Ong, Yen Chin",
    title = "{Insights in f(Q) cosmology: the relevance of the connection}",
    eprint = "2506.03506",
    archivePrefix = "arXiv",
    primaryClass = "gr-qc",
    reportNumber = "RIKEN-iTHEMS-Report-25",
    doi = "10.1088/1475-7516/2025/11/068",
    journal = "JCAP",
    volume = "11",
    pages = "068",
    year = "2025"
}

@article{Akarsu:2021fol,
    author = {Akarsu, \"Ozg\"ur and Kumar, Suresh and \"Oz\"ulker, Emre and Vazquez, J. Alberto},
    title = "{Relaxing cosmological tensions with a sign switching cosmological constant}",
    eprint = "2108.09239",
    archivePrefix = "arXiv",
    primaryClass = "astro-ph.CO",
    doi = "10.1103/PhysRevD.104.123512",
    journal = "Phys. Rev. D",
    volume = "104",
    number = "12",
    pages = "123512",
    year = "2021"
}

@article{Aurilia:1980xj,
    author = "Aurilia, Antonio and Nicolai, H. and Townsend, P. K.",
    title = "{Hidden Constants: The Theta Parameter of QCD and the Cosmological Constant of N=8 Supergravity}",
    reportNumber = "CERN-TH-2875",
    doi = "10.1016/0550-3213(80)90466-6",
    journal = "Nucl. Phys. B",
    volume = "176",
    pages = "509--522",
    year = "1980"
}

@article{Duff:1980qv,
    author = "Duff, M. J. and van Nieuwenhuizen, P.",
    title = "{Quantum Inequivalence of Different Field Representations}",
    reportNumber = "ICTP/79-80/33",
    doi = "10.1016/0370-2693(80)90852-7",
    journal = "Phys. Lett. B",
    volume = "94",
    pages = "179--182",
    year = "1980"
}

@article{Koivisto:2009ew,
    author = "Koivisto, Tomi S. and Nunes, Nelson J.",
    title = "{Three-form cosmology}",
    eprint = "0907.3883",
    archivePrefix = "arXiv",
    primaryClass = "astro-ph.CO",
    doi = "10.1016/j.physletb.2010.01.051",
    journal = "Phys. Lett. B",
    volume = "685",
    pages = "105--109",
    year = "2010"
}

@article{DeFelice:2012jt,
    author = "De Felice, Antonio and Karwan, Khamphee and Wongjun, Pitayuth",
    title = "{Stability of the 3-form field during inflation}",
    eprint = "1202.0896",
    archivePrefix = "arXiv",
    primaryClass = "hep-ph",
    doi = "10.1103/PhysRevD.85.123545",
    journal = "Phys. Rev. D",
    volume = "85",
    pages = "123545",
    year = "2012"
}

@article{DeFelice:2012wy,
    author = "De Felice, Antonio and Karwan, Khamphee and Wongjun, Pitayuth",
    title = "{Reheating in 3-form inflation}",
    eprint = "1209.5156",
    archivePrefix = "arXiv",
    primaryClass = "astro-ph.CO",
    doi = "10.1103/PhysRevD.86.103526",
    journal = "Phys. Rev. D",
    volume = "86",
    pages = "103526",
    year = "2012"
}

@article{Mulryne_2012,
   title={Three-form inflation and non-Gaussianity},
   volume={2012},
   ISSN={1475-7516},
   url={http://dx.doi.org/10.1088/1475-7516/2012/12/016},
   DOI={10.1088/1475-7516/2012/12/016},
   number={12},
   journal={Journal of Cosmology and Astroparticle Physics},
   publisher={IOP Publishing},
   author={Mulryne, David J and Noller, Johannes and Nunes, Nelson J},
   year={2012},
   month=dec, pages={016–016} }

@article{Kumar:2014oka,
    author = "Kumar, K. Sravan and Marto, J. and Nunes, Nelson J. and Moniz, P. Vargas",
    title = "{Inflation in a two 3-form fields scenario}",
    eprint = "1404.0211",
    archivePrefix = "arXiv",
    primaryClass = "gr-qc",
    doi = "10.1088/1475-7516/2014/06/064",
    journal = "JCAP",
    volume = "06",
    pages = "064",
    year = "2014"
}

@article{SravanKumar:2016biw,
    author = "Sravan Kumar, K. and Mulryne, David J. and Nunes, Nelson J. and Marto, Jo{\~a}o and Vargas Moniz, Paulo",
    title = "{Non-Gaussianity in multiple three-form field inflation}",
    eprint = "1606.07114",
    archivePrefix = "arXiv",
    primaryClass = "astro-ph.CO",
    doi = "10.1103/PhysRevD.94.103504",
    journal = "Phys. Rev. D",
    volume = "94",
    number = "10",
    pages = "103504",
    year = "2016"
}

@article{Barros:2015evi,
    author = "Barros, Bruno J. and Nunes, Nelson J.",
    title = "{Three-form inflation in type II Randall-Sundrum}",
    eprint = "1511.07856",
    archivePrefix = "arXiv",
    primaryClass = "astro-ph.CO",
    doi = "10.1103/PhysRevD.93.043512",
    journal = "Phys. Rev. D",
    volume = "93",
    number = "4",
    pages = "043512",
    year = "2016"
}

@article{Bouhmadi-Lopez:2021zwt,
    author = "Bouhmadi-L{\'o}pez, Mariam and Chen, Che-Yu and Chew, Xiao Yan and Ong, Yen Chin and Yeom, Dong-han",
    title = "{Traversable wormhole in Einstein 3-form theory with self-interacting potential}",
    eprint = "2108.07302",
    archivePrefix = "arXiv",
    primaryClass = "gr-qc",
    doi = "10.1088/1475-7516/2021/10/059",
    journal = "JCAP",
    volume = "10",
    pages = "059",
    year = "2021"
}

@article{Bouhmadi-Lopez:2020wve,
    author = "Bouhmadi-L{\'o}pez, Mariam and Chen, Che-Yu and Chew, Xiao Yan and Ong, Yen Chin and Yeom, Dong-Han",
    title = "{Regular Black Hole Interior Spacetime Supported by Three-Form Field}",
    eprint = "2005.13260",
    archivePrefix = "arXiv",
    primaryClass = "gr-qc",
    doi = "10.1140/epjc/s10052-021-09080-1",
    journal = "Eur. Phys. J. C",
    volume = "81",
    number = "4",
    pages = "278",
    year = "2021"
}

@article{Barros:2020ghz,
    author = "Barros, Bruno J. and Dǎnilǎ, Bogdan and Harko, Tiberiu and Lobo, Francisco S. N.",
    title = "{Black hole and naked singularity geometries supported by three-form fields}",
    eprint = "2004.06605",
    archivePrefix = "arXiv",
    primaryClass = "gr-qc",
    doi = "10.1140/epjc/s10052-020-8178-1",
    journal = "Eur. Phys. J. C",
    volume = "80",
    pages = "617",
    year = "2020"
}

@article{daFonseca:2024boz,
    author = "da Fonseca, Vitor and Barros, Bruno J. and Barreiro, Tiago and Nunes, Nelson J.",
    title = "{Non-canonical 3-form dark energy}",
    eprint = "2410.11658",
    archivePrefix = "arXiv",
    primaryClass = "gr-qc",
    doi = "10.1016/j.dark.2025.101827",
    journal = "Phys. Dark Univ.",
    volume = "47",
    pages = "101827",
    year = "2025"
}

@article{DeFelice:2025khe,
    author = "De Felice, Antonio and Hell, Anamaria",
    title = "{The non-minimal 3-form cosmology and the rise of the cuscuton}",
    eprint = "2509.02323",
    archivePrefix = "arXiv",
    primaryClass = "gr-qc",
    reportNumber = "IPMU25-0044, YITP-25-136",
    doi = "10.1007/JHEP11(2025)132",
    journal = "JHEP",
    volume = "11",
    pages = "132",
    year = "2025"
}

@article{Bouhmadi-Lopez:2018lly,
    author = "Bouhmadi-L{\'o}pez, Mariam and Brizuela, David and Garay, I{\~n}aki",
    title = "{Quantum behavior of the ''Little Sibling'' of the Big Rip induced by a three-form field}",
    eprint = "1802.05164",
    archivePrefix = "arXiv",
    primaryClass = "gr-qc",
    doi = "10.1088/1475-7516/2018/09/031",
    journal = "JCAP",
    volume = "09",
    pages = "031",
    year = "2018"
}

@article{Barros:2023nzr,
    author = "Barros, Bruno J. and Beltr{\'a}n Jim{\'e}nez, Jose",
    title = "{Non-trivial thick brane realisations with 3-forms}",
    eprint = "2312.12516",
    archivePrefix = "arXiv",
    primaryClass = "gr-qc",
    doi = "10.1007/JHEP02(2024)002",
    journal = "JHEP",
    volume = "02",
    pages = "002",
    year = "2024"
}

@article{Koivisto_2013,
   title={Coupled three-form dark energy},
   volume={88},
   ISSN={1550-2368},
   url={http://dx.doi.org/10.1103/PhysRevD.88.123512},
   DOI={10.1103/physrevd.88.123512},
   number={12},
   journal={Physical Review D},
   publisher={American Physical Society (APS)},
   author={Koivisto, Tomi S. and Nunes, Nelson J.},
   year={2013},
   month=dec }

@article{Barros_2018,
   title={Wormhole geometries supported by three-form fields},
   volume={98},
   ISSN={2470-0029},
   url={http://dx.doi.org/10.1103/PhysRevD.98.044012},
   DOI={10.1103/physrevd.98.044012},
   number={4},
   journal={Physical Review D},
   publisher={American Physical Society (APS)},
   author={Barros, Bruno J. and Lobo, Francisco S. N.},
   year={2018},
   month=aug }

@article{Barreiro:2016aln,
    author = "Barreiro, Tiago and Bertello, Ugo and Nunes, Nelson J.",
    title = "{Screening three-form fields}",
    eprint = "1610.00357",
    archivePrefix = "arXiv",
    primaryClass = "gr-qc",
    doi = "10.1016/j.physletb.2017.08.061",
    journal = "Phys. Lett. B",
    volume = "773",
    pages = "417--421",
    year = "2017"
}

@article{Akarsu:2022typ,
    author = {Akarsu, Ozgur and Kumar, Suresh and \"Oz\"ulker, Emre and Vazquez, J. Alberto and Yadav, Anita},
    title = "{Relaxing cosmological tensions with a sign switching cosmological constant: Improved results with Planck, BAO, and Pantheon data}",
    eprint = "2211.05742",
    archivePrefix = "arXiv",
    primaryClass = "astro-ph.CO",
    doi = "10.1103/PhysRevD.108.023513",
    journal = "Phys. Rev. D",
    volume = "108",
    number = "2",
    pages = "023513",
    year = "2023"
}

@article{Urban:2012ib,
    author = "Urban, Federico R. and Koivisto, Tomi S.",
    title = "{Perturbations and non-Gaussianities in three-form inflationary magnetogenesis}",
    eprint = "1207.7328",
    archivePrefix = "arXiv",
    primaryClass = "astro-ph.CO",
    doi = "10.1088/1475-7516/2012/09/025",
    journal = "JCAP",
    volume = "09",
    pages = "025",
    year = "2012"
}

@article{Gordin:2023nsv,
    author = "Gordin, Jake E. B. and MacDevette, Kelly and Bruton, Jenna",
    title = "{The dynamics of three-forms in thick branes}",
    eprint = "2311.14436",
    archivePrefix = "arXiv",
    primaryClass = "hep-th",
    doi = "10.1007/JHEP05(2024)061",
    journal = "JHEP",
    volume = "05",
    pages = "061",
    year = "2024"
}

@article{Morais:2017vlf,
    author = "Morais, Jo{\~a}o and Bouhmadi-L{\'o}pez, Mariam and Marto, Jo{\~a}o",
    editor = {D{\k{a}}browski, Mariusz P. and Kr{\"a}mer, Manuel and Salzano, Vincenzo},
    title = "{3-Form Cosmology: Phantom Behaviour, Singularities and Interactions}",
    doi = "10.3390/universe3010021",
    journal = "Universe",
    volume = "3",
    number = "1",
    pages = "21",
    year = "2017"
}

@article{Barros:2021jbt,
    author = "Barros, Bruno J. and Haghani, Zahra and Harko, Tiberiu and Lobo, Francisco S. N.",
    title = "{Static spherically symmetric three-form stars}",
    eprint = "2101.04445",
    archivePrefix = "arXiv",
    primaryClass = "gr-qc",
    doi = "10.1140/epjc/s10052-021-09105-9",
    journal = "Eur. Phys. J. C",
    volume = "81",
    number = "4",
    pages = "307",
    year = "2021"
}

@article{Akarsu:2023mfb,
    author = "Akarsu, Ozgur and Di Valentino, Eleonora and Kumar, Suresh and Nunes, Rafael C. and Vazquez, J. Alberto and Yadav, Anita",
    title = "{$\Lambda_{\rm s}$CDM model: A promising scenario for alleviation of cosmological tensions}",
    eprint = "2307.10899",
    archivePrefix = "arXiv",
    primaryClass = "astro-ph.CO",
    month = "7",
    year = "2023"
}

@article{Bouhmadi-Lopez:2025ggl,
    author = "Bouhmadi-L{\'o}pez, Mariam and Ibarra-Uriondo, Be{\~n}at",
    title = "{Cosmographic analysis of sign-switching dark energy}",
    eprint = "2506.12139",
    archivePrefix = "arXiv",
    primaryClass = "gr-qc",
    doi = "10.1103/v1cl-pr54",
    journal = "Phys. Rev. D",
    volume = "112",
    number = "6",
    pages = "063559",
    year = "2025"
}

@article{Bouhmadi-Lopez:2025spo,
    author = "Bouhmadi-L{\'o}pez, Mariam and Ibarra-Uriondo, Be{\~n}at",
    title = "{Cosmological perturbations for smooth sign-switching dark energy models}",
    eprint = "2506.18992",
    archivePrefix = "arXiv",
    primaryClass = "gr-qc",
    doi = "10.1016/j.dark.2025.102129",
    journal = "Phys. Dark Univ.",
    volume = "50",
    pages = "102129",
    year = "2025"
}

@article{Germani:2009iq,
    author = "Germani, Cristiano and Kehagias, Alex",
    title = "{P-nflation: generating cosmic Inflation with p-forms}",
    eprint = "0902.3667",
    archivePrefix = "arXiv",
    primaryClass = "astro-ph.CO",
    doi = "10.1088/1475-7516/2009/03/028",
    journal = "JCAP",
    volume = "03",
    pages = "028",
    year = "2009"
}

@article{DiValentino2021,
    author = {Di Valentino, E. and others},
    title = {Cosmology Intertwined: A Review of the Particle Physics, Astrophysics, and Cosmology Associated with the Cosmological Tensions and Anomalies},
    journal = {Journal of High Energy Astrophysics},
    volume = {30},
    pages = {49},
    year = {2021},
    doi = {10.1016/j.jheap.2021.03.001},
}

@article{Boiza:2024azh,
    author = "Boiza, Carlos G. and Bouhmadi-L{\'o}pez, Mariam",
    title = "{Speeding up the Universe with a generalised axion-like potential}",
    eprint = "2409.18184",
    archivePrefix = "arXiv",
    primaryClass = "astro-ph.CO",
    doi = "10.1140/epjc/s10052-025-14490-6",
    journal = "Eur. Phys. J. C",
    volume = "85",
    number = "7",
    pages = "777",
    year = "2025"
}

@article{Torrado:2020dgo,
    author = "Torrado, Jesus and Lewis, Antony",
    title = "{Cobaya: Code for Bayesian Analysis of hierarchical physical models}",
    eprint = "2005.05290",
    archivePrefix = "arXiv",
    primaryClass = "astro-ph.IM",
    reportNumber = "TTK-20-15",
    doi = "10.1088/1475-7516/2021/05/057",
    journal = "JCAP",
    volume = "05",
    pages = "057",
    year = "2021"
}

@article{2019ascl.soft10019T,
       author = {{Torrado}, Jes{\'u}s and {Lewis}, Antony},
        title = "{Cobaya: Bayesian analysis in cosmology}",
 howpublished = {Astrophysics Source Code Library, record ascl:1910.019},
         year = 2019,
        month = oct,
          eid = {ascl:1910.019},
archivePrefix = {ascl},
       eprint = {1910.019},
       adsurl = {https://ui.adsabs.harvard.edu/abs/2019ascl.soft10019T}
}

@article{Lewis:2002ah,
    author = "Lewis, Antony and Bridle, Sarah",
    title = "{Cosmological parameters from CMB and other data: A Monte Carlo approach}",
    eprint = "astro-ph/0205436",
    archivePrefix = "arXiv",
    doi = "10.1103/PhysRevD.66.103511",
    journal = "Phys. Rev. D",
    volume = "66",
    pages = "103511",
    year = "2002"
}

@article{neal2005takingbiggermetropolissteps,
      title={Taking Bigger Metropolis Steps by Dragging Fast Variables}, 
      author={Radford M. Neal},
      year={2005},
      eprint={math/0502099},
      archivePrefix={arXiv},
      primaryClass={math.ST},
      url={https://arxiv.org/abs/math/0502099}, 
}

@article{Lewis:2013hha,
    author = "Lewis, Antony",
    title = "{Efficient sampling of fast and slow cosmological parameters}",
    eprint = "1304.4473",
    archivePrefix = "arXiv",
    primaryClass = "astro-ph.CO",
    doi = "10.1103/PhysRevD.87.103529",
    journal = "Phys. Rev. D",
    volume = "87",
    number = "10",
    pages = "103529",
    year = "2013"
}

@article{Hu:1995en,
    author = "Hu, Wayne and Sugiyama, Naoshi",
    title = "{Small scale cosmological perturbations: An Analytic approach}",
    eprint = "astro-ph/9510117",
    archivePrefix = "arXiv",
    reportNumber = "IASSNS-AST-95-42, CFPA-TH-95-18, UTAP-212",
    doi = "10.1086/177989",
    journal = "Astrophys. J.",
    volume = "471",
    pages = "542--570",
    year = "1996"
}

@article{Eisenstein:1997ik,
    author = "Eisenstein, Daniel J. and Hu, Wayne",
    title = "{Baryonic features in the matter transfer function}",
    eprint = "astro-ph/9709112",
    archivePrefix = "arXiv",
    reportNumber = "IASSNS-AST-97-51",
    doi = "10.1086/305424",
    journal = "Astrophys. J.",
    volume = "496",
    pages = "605",
    year = "1998"
}

@article{Fixsen:2009ug,
    author = "Fixsen, D. J.",
    title = "{The Temperature of the Cosmic Microwave Background}",
    eprint = "0911.1955",
    archivePrefix = "arXiv",
    primaryClass = "astro-ph.CO",
    doi = "10.1088/0004-637X/707/2/916",
    journal = "Astrophys. J.",
    volume = "707",
    pages = "916--920",
    year = "2009"
}

@article{Wang_2006,
   title={Robust Dark Energy Constraints from Supernovae, Galaxy Clustering, and 3 yrWilkinson Microwave Anisotropy ProbeObservations},
   volume={650},
   ISSN={1538-4357},
   url={http://dx.doi.org/10.1086/507091},
   DOI={10.1086/507091},
   number={1},
   journal={The Astrophysical Journal},
   publisher={American Astronomical Society},
   author={Wang, Yun and Mukherjee, Pia},
   year={2006},
   month=oct, pages={1–6} }

@article{Abdul_Karim_2025i,
   title={DESI DR2 results. I. Baryon acoustic oscillations from the Lyman alpha forest},
   volume={112},
   ISSN={2470-0029},
   url={http://dx.doi.org/10.1103/2wwn-xjm5},
   DOI={10.1103/2wwn-xjm5},
   number={8},
   journal={Physical Review D},
   publisher={American Physical Society (APS)},
   eprint = "2503.14739",
   archivePrefix = "arXiv",
   author = "Abdul Karim, M. and others",
   collaboration="DESI Collaboration",
   year={2025},
   month=oct }

@article{Abdul_Karim_2025ii,
   title={DESI DR2 results. II. Measurements of baryon acoustic oscillations and cosmological constraints},
   volume={112},
   ISSN={2470-0029},
   url={http://dx.doi.org/10.1103/tr6y-kpc6},
   DOI={10.1103/tr6y-kpc6},
   number={8},
   journal={Physical Review D},
   publisher={American Physical Society (APS)},
   eprint = "2503.14738",
   archivePrefix = "arXiv",
   author = "Abdul Karim, M. and others",
   collaboration = "DESI Collaboration",
   year={2025},
   month=oct }

@article{Brieden_2023,
   title={A tale of two (or more) h’s},
   volume={2023},
   ISSN={1475-7516},
   url={http://dx.doi.org/10.1088/1475-7516/2023/04/023},
   DOI={10.1088/1475-7516/2023/04/023},
   number={04},
   journal={Journal of Cosmology and Astroparticle Physics},
   publisher={IOP Publishing},
   author={Brieden, Samuel and Gil-Marín, Héctor and Verde, Licia},
   year={2023},
   month=apr, pages={023} }

@article{Pan+S,
   title={The Pantheon+ Analysis: Cosmological Constraints},
   volume={938},
   ISSN={1538-4357},
   url={http://dx.doi.org/10.3847/1538-4357/ac8e04},
   DOI={10.3847/1538-4357/ac8e04},
   number={2},
   journal={The Astrophysical Journal},
   publisher={American Astronomical Society},
   eprint = "2202.04077",
   archivePrefix = "arXiv",
   author={Brout et al.},
   year={2022},
   month=oct, pages={110} }

@article{Zhai_2019,
   title={Robust and model-independent cosmological constraints from distance measurements},
   volume={2019},
   ISSN={1475-7516},
   url={http://dx.doi.org/10.1088/1475-7516/2019/07/005},
   DOI={10.1088/1475-7516/2019/07/005},
   number={07},
   journal={Journal of Cosmology and Astroparticle Physics},
   publisher={IOP Publishing},
   author={Zhai, Zhongxu and Wang, Yun},
   year={2019},
   month=jul, pages={005–005} }

@article{GetDist,
   title={GetDist: a Python package for analysing Monte Carlo samples},
   volume={2025},
   ISSN={1475-7516},
   url={http://dx.doi.org/10.1088/1475-7516/2025/08/025},
   DOI={10.1088/1475-7516/2025/08/025},
   number={08},
   journal={Journal of Cosmology and Astroparticle Physics},
   publisher={IOP Publishing},
   author={Lewis, Antony},
   year={2025},
   month=aug, pages={025} }

@article{Schoneberg:2024ifp,
    author = {Sch{\"o}neberg, Nils},
    title = "{The 2024 BBN baryon abundance update}",
    eprint = "2401.15054",
    archivePrefix = "arXiv",
    primaryClass = "astro-ph.CO",
    doi = "10.1088/1475-7516/2024/06/006",
    journal = "JCAP",
    volume = "06",
    pages = "006",
    year = "2024"
}

@article{Jimenez:2001gg,
    author = "Jimenez, Raul and Loeb, Abraham",
    title = "{Constraining cosmological parameters based on relative galaxy ages}",
    eprint = "astro-ph/0106145",
    archivePrefix = "arXiv",
    doi = "10.1086/340549",
    journal = "Astrophys. J.",
    volume = "573",
    pages = "37--42",
    year = "2002"
}

@article{Moresco:2024wmr,
    author = "Moresco, Michele",
    title = "{Measuring the expansion history of the Universe with cosmic chronometers}",
    eprint = "2412.01994",
    archivePrefix = "arXiv",
    primaryClass = "astro-ph.CO",
    month = "12",
    year = "2024"
}

@article{Liu:2014vda,
    author = "Liu, Jing and Wei, Hao",
    title = "{Cosmological models and gamma-ray bursts calibrated by using Pad{\'e} method}",
    eprint = "1410.3960",
    archivePrefix = "arXiv",
    primaryClass = "astro-ph.CO",
    doi = "10.1007/s10714-015-1986-1",
    journal = "Gen. Rel. Grav.",
    volume = "47",
    number = "11",
    pages = "141",
    year = "2015"
}

@article{Conley_2010,
   title={SUPERNOVA CONSTRAINTS AND SYSTEMATIC UNCERTAINTIES FROM THE FIRST THREE YEARS OF THE SUPERNOVA LEGACY SURVEY},
   volume={192},
   ISSN={1538-4365},
   url={http://dx.doi.org/10.1088/0067-0049/192/1/1},
   DOI={10.1088/0067-0049/192/1/1},
   eprint = "1104.1443",
   archivePrefix = "arXiv",
   number={1},
   journal={The Astrophysical Journal Supplement Series},
   publisher={American Astronomical Society},
   author={Conley, A. and others},
   year={2010},
   month=dec, pages={1} }

@article{Liddle_2007,
   title={Information criteria for astrophysical model selection},
   volume={377},
   ISSN={1745-3925},
   url={http://dx.doi.org/10.1111/j.1745-3933.2007.00306.x},
   DOI={10.1111/j.1745-3933.2007.00306.x},
   number={1},
   journal={Monthly Notices of the Royal Astronomical Society: Letters},
   publisher={Oxford University Press (OUP)},
   author={Liddle, Andrew R.},
   year={2007},
   month=may, pages={L74–L78},
   eprint = "astro-ph/0701113",
   archivePrefix = "arXiv",}

@book{Kenneth:2004a,
    author = "Burnham, Kenneth and Anderson, David",
    title = "{Model Selection and Multimodel Inference: A Practical Information-theoretic Approach}",
    edition = "2nd",
    publisher = "Springer", 
    address = "New York",
    year = "2002"
}

@article{Kenneth:2004b,
    author = "Kenneth P. Burnham and David R. Anderson",
    title ="{Multimodel Inference: Understanding AIC and BIC in Model Selection}",
    journal = "Sociological Methods \& Research",
    volume = "33",
    number = "2",
    pages = "261-304",
    year = "2004",
    doi = "10.1177/0049124104268644",
}

@article{Lobo:2026flk,
    author = "Lobo, Francisco S. N. and Rodrigues, Manuel E.",
    title = "{Black Bounce Solutions from a Self-Interacting 3-Form Field in General Relativity}",
    eprint = "2606.05508",
    archivePrefix = "arXiv",
    primaryClass = "gr-qc",
    month = "6",
    year = "2026"
}

@article{Bouhmadi-Lopez:2026vyc,
    author = "Bouhmadi-L{\'o}pez, Mariam and Chiang, Hsu-Wen and Ibarra-Uriondo, Be{\~n}at",
    title = "{Alleviating the Hubble Tension with Smooth Sign-Switching Dark Energy: Full CMB Constraints with DESI and PantheonPlus}",
    eprint = "2607.05044",
    archivePrefix = "arXiv",
    primaryClass = "astro-ph.CO",
    month = "7",
    year = "2026"
}

@article{Klinkhamer:2008ns,
    author = "Klinkhamer, F. R. and Volovik, G. E.",
    title = "{Dynamic vacuum variable and equilibrium approach in cosmology}",
    eprint = "0806.2805",
    archivePrefix = "arXiv",
    primaryClass = "gr-qc",
    reportNumber = "KA-TP-14-2008",
    doi = "10.1103/PhysRevD.78.063528",
    journal = "Phys. Rev. D",
    volume = "78",
    pages = "063528",
    year = "2008"
}

@article{Klinkhamer:2007pe,
    author = "Klinkhamer, F. R. and Volovik, G. E.",
    title = "{Self-tuning vacuum variable and cosmological constant}",
    eprint = "0711.3170",
    archivePrefix = "arXiv",
    primaryClass = "gr-qc",
    reportNumber = "KA-TP-29-2007",
    doi = "10.1103/PhysRevD.77.085015",
    journal = "Phys. Rev. D",
    volume = "77",
    pages = "085015",
    year = "2008"
}

@article{Cai:2026swf,
  author = {Cai, Rong-Gen and Wang, Shao-Jiang},
  title = {The {Hubble} tension: a decade review},
  journal = {Res. Astron. Astrophys.},
  volume = {26},
  pages = {084011},
  year = {2026},
  doi = {10.1088/1674-4527/ae842f},
  eprint = {2606.20434},
  archivePrefix = {arXiv},
  primaryClass = {astro-ph.CO}
}

\end{document}